\documentclass[11pt]{article}
\newcommand{\Comment}[1]{{}}
\usepackage{amsfonts,amsthm,amsmath,amssymb,slashed}
\usepackage[textwidth = 480 pt, textheight = 650 pt]{geometry}

\usepackage{color}
\definecolor{MyDarkBlue}{rgb}{0.15,0.15,0.45}
\usepackage[linktocpage=true]{hyperref}
\hypersetup{
colorlinks=true,
citecolor=MyDarkBlue,
linkcolor=MyDarkBlue,
urlcolor=MyDarkBlue,
pdfauthor={Horatiu Nastase and Constantinos Papageorgakis},
pdftitle={Fuzzy Killing Spinors and Supersymmetric D4 action on the Fuzzy 2-sphere from the ABJM Model},
pdfsubject={hep-th}
}

\usepackage[numbers,sort&compress]{natbib}
\usepackage{hypernat}

\newcommand\sfrac[2]{{\textstyle\frac{#1}{#2}}}
\newcommand\ignore[1]{}
\def\one{{\,\hbox{1\kern-.8mm l}}}

\def\Tr{{\rm Tr\, }}

\newcommand{\SO}{\mathrm{SO}} 
\newcommand{\SU}{\mathrm{SU}} \newcommand{\U}{\mathrm{U}}
\newcommand{\ie}{\emph{i.e.}\:}
\newcommand{\eg}{\emph{e.g.}\;} \newcommand{\pd}{\partial}
\newcommand{\doublet}[2]{\left(\begin{array}{c}#1\\#2\end{array}\right)}
\newcommand{\twobytwo}[4]{\left(\begin{array}{cc} #1&#2\\#3&#4\end{array}\right)}

\def\a{\alpha}\def\b{\beta}
\def\g{\gamma}

\def\s{\sigma}

\def\d{\partial}

\newcommand{\cp}{\Cset \mathrm{P}}
\newcommand{\Cset}{{\,\,{{{^{_{\pmb{\mid}}}}\kern-.45em{\mathrm C}}}}}

\newcommand{\cA}{\mathcal A}

\newcommand{\cN}{\mathcal N}

\newcommand{\be}{\begin{equation}}
\newcommand{\bea}{\begin{eqnarray}}

\newcommand{\ee}{\end{equation}}
\newcommand{\eea}{\end{eqnarray}}

\newcommand{\nn}{\nonumber}

\def\ts{ \tilde \sigma }

\begin{document}

\makeatletter
\@addtoreset{equation}{section}
\makeatother
\renewcommand{\theequation}{\thesection.\arabic{equation}}

\rightline{TIT/HEP-598}
\rightline{TIFR/TH/09-21}
   \vspace{1.8truecm}

\vspace{15pt}

%%%%%%%%%%%%%%%%%

\centerline{\LARGE \bf Fuzzy Killing Spinors and Supersymmetric D4 action   } 
\centerline{ \LARGE\bf  on the Fuzzy 2-sphere from the ABJM Model} \vspace{1truecm}
\thispagestyle{empty} \centerline{
    {\large \bf Horatiu Nastase${}^{a,}$}\footnote{E-mail address: \href{mailto:nastase.h.aa@m.titech.ac.jp}{\tt nastase.h.aa@m.titech.ac.jp}} {\bf and}
    {\large \bf Constantinos Papageorgakis${}^{b,}$}\footnote{E-mail address:
                                 \href{mailto:costis@theory.tifr.res.in}{\tt costis@theory.tifr.res.in}}
                                                           }

\vspace{1cm}
\centerline{{\it ${}^a$ Global Edge Institute, Tokyo Institute of Technology,}}
\centerline{{\it Meguro, Tokyo 152-8550, Japan}}

\vspace{.4cm}
\centerline{{\it ${}^b$ 
 Tata
    Institute of Fundamental Research},} \centerline{{\it Homi Bhabha
    Road, Mumbai 400 005, India}}

\vspace{1.4truecm}

%%%%%%%%%%%%%%%%%
\thispagestyle{empty}

\centerline{\bf ABSTRACT}

\vspace{.4truecm}

\noindent
Our recent construction arXiv:0903.3966 for the fuzzy 2-sphere in terms of bifundamentals, discovered in the context of the ABJM model, is shown to be explicitly equivalent to the usual (adjoint) fuzzy sphere construction.  The matrices $\tilde{G}^\a$ that define it play the role of fuzzy Killing spinors on the 2-sphere, out of which all spherical harmonics are constructed. Starting from the quadratic fluctuation action around these solutions in the mass-deformed ABJM theory, we recover a supersymmetric D4-brane action wrapping a 2-sphere, including fermions. We obtain both the usual D4 action with an unusual $x$-dependence on the sphere, as well as a twisted version in terms of the usual $x$-dependence, and contrast our result with the Maldacena-N\'u\~nez case of a D5 wrapping an $S^2$. The twisted and unwisted fields are related by the same matrix $\tilde{G}^\a$.

\vspace{.5cm}

\setcounter{page}{0}
\setcounter{tocdepth}{2}

\newpage

\tableofcontents

\setcounter{footnote}{0}

\linespread{1.1}
\parskip 4pt

{}~
{}~

\section{Introduction}\label{intro}

The construction of worldvolume theories potentially capturing the dynamics of multiple M2-branes have recently received much attention.  Motivated in part by the proposed BPS equation for M2$\perp$M5 systems of Basu-Harvey \cite{Basu:2004ed} (corresponding to a `fuzzy funnel') and the work of Schwarz \cite{Schwarz:2004yj}, Bagger-Lambert and independently Gustavsson (BLG) \cite{Bagger:2006sk,Bagger:2007jr,Bagger:2007vi,Gustavsson:2007vu} proposed a maximally (${\cal N}=8$) supersymmetric theory in 2+1 dimensions based on a general 3-algebra. It was subsequently proved that for positive definite 3-algebra metric, the only possibility is the four-dimensional 3-algebra or ${\cal A}_4$-theory \cite{Papadopoulos:2008sk,Gauntlett:2008uf}, which was reformulated by Van Raamsdonk in terms of a conventional Chern-Simons (CS) field theory with gauge group $\SU(2)_k\times \SU(2)_{-k}$ and bifundamental matter fields \cite{VanRaamsdonk:2008ft}. Using the novel Higgs mechanism of \cite{Mukhi:2008ux}\footnote{See also \cite{Nicolai:2003bp} for   earlier work.} it was initially conjectured that this theory describes 2 M2-branes on some exotic orbifold of M-theory \cite{Lambert:2008et,Distler:2008mk}. Subsequently, and also following developments in supersymmetric 2+1d CS theories \cite{Gaiotto:2008sd,Hosomichi:2008jd}, Aharony, Bergman, Jafferis and Maldacena (ABJM) \cite{Aharony:2008ug} concretely realised the above ideas in the form of a $\U(N)_k\times \U(N)_{-k}$ Chern-Simons gauge theory with bifundamentals as the low-energy limit for the theory of $N$ M2-branes living on a ${\mathbb C}^4/{\mathbb Z}_k$ transverse space, with the trade-off of reducing the supersymmetry from ${\cal N}=8$ to ${\cal N}=6$. In the case of CS level $k=1,2$ the supersymmetry is expected to get enhanced back to ${\cal N}=8$ and this has been recently confirmed explicitly with the use of monopole operators \cite{Benna:2009xd,Gustavsson:2009pm,Kwon:2009ar}. The implications of the ABJM model in the context of $\text{AdS}_4/ \text{CFT}_3$ generated great interest, that however we will not review here. The theory was extended to an $\U(N)\times \U(M)$ gauge group in \cite{Aharony:2008gk,Hosomichi:2008jb}. Based on the maximally supersymmetric massive deformation of BLG defined in \cite{Bagger:2007vi,Gomis:2008cv,Hosomichi:2008jb}, a maximally supersymmetric massive deformation of ABJM was given in \cite{Gomis:2008vc} by Gomis, Rodr\'iguez-G\'omez, Van Raamsdonk and Verlinde (GRVV), the ground state of which is a fuzzy sphere solution expected to describe a configuration of M2-branes blowing up into spherical M5-branes through the Myers effect \cite{Myers:1999ps}. At $k=1$ these solutions should have a dual description in terms of the $\sfrac{1}{2}$-BPS M-theory geometries with flux found in \cite{Bena:2004jw,Lin:2004nb}.

The original motivation for the latter developments, \ie the hope of explicitly describing a BPS fuzzy 3-funnel for M2$\perp$M5 systems in the pure (undeformed) ABJM theory, or the fuzzy 3-sphere for M2-M5 bound states in the case of the massive deformation, turned out to be unfounded. Indeed, in \cite{Nastase:2009ny} it was shown that, at least in perturbation theory where the ABJM coupling $\lambda = \frac{N}{k}$ is small, the particular gauge theory solution for $N>2$ (and thus also in the classical, large-$N$ limit) `deconstructs' a fuzzy 2-sphere instead of a fuzzy 3-sphere.\footnote{Note that this means that at finite $k>N>2$, only nonperturbative   effects could turn the fuzzy 2-sphere into a fuzzy 3-sphere, and it is difficult to   see how that can happen.}  This is a natural expectation, as these configurations were found to sport only an $\SU(2)$ symmetry, to be contrasted with the $\SO(4)$-covariant fuzzy 3-sphere construction of Guralnik-Ramgoolam \cite{Guralnik:2000pb,   Ramgoolam:2001zx,Ramgoolam:2002wb}. Furthermore, the bosonic part of the action for small fluctuations was also found to be consistent with a fuzzy 2-sphere. Only when $N=2$, the case of the `fuzziest' (most quantum) sphere, corresponding to the ${\cal A}_4$ BLG model, is the $\SO(4)$ symmetry manifest and thus the solution could be interpreted as a (very fuzzy) 3-sphere. Unfortunately,  no classical limit is possible in that example, as $N$ is fixed.

The above results are obtained at finite $N,k$. For the classical (large-$N$) limit, the brane interpretation, as given in \cite{Nastase:2009ny}, is that in perturbation theory one is forced by the classical (large-$N$) limit to also consider $k$ large. Then, one descends from M-theory down to type IIA, as the M5 wrapping an $S^3$ modded out by the ${\mathbb Z}_k$ action of the ${\mathbb C}^4/{\mathbb Z}_k$ target space is reduced to a D4 wrapping an $S^2$; this is realised as the $S^1/{\mathbb   Z}_k\hookrightarrow S^3/{\mathbb Z}_k\stackrel{\pi}{\rightarrow} S^2$ Hopf fibration, with the $S^1$ fibre shrinking $k$ times, and hence to zero in the $k\rightarrow\infty$ limit.  Note in the $k=1,2$ case it should be possible to take the classical large $N$ limit in a different way so that a 3-sphere does emerge, but this would be a construction for solutions carrying nonperturbative charges, like \eg the ones studied recently in \cite{Auzzi:2009es}, which we currently lack the tools to analyse.  Conceivably, the presence of the monopole operators, which enhance supersymmetry from ${\cal N}=6$ to ${\cal N}=8$, could also enhance the fuzzy sphere symmetry to $\SO(4)$. For $k>2$ and fixed however, it is hard to see how nonperturbative effects could change the symmetry of the fuzzy sphere.

The fuzzy 2-sphere construction that appears in the above systems, as the (fuzzy) base of the Hopf fibration, is an interesting new model emerging out of \emph{bifundamental} instead of the usual adjoint matrices. It should have wider applicability, in the context of general quiver gauge theories with bifundamental matter that admit fuzzy sphere solutions, as can also be seen in \cite{Maldacena:2009mw}.

In this work we continue the study of this bifundamental fuzzy 2-sphere realisation and its role in the ABJM model.  We first set out to understand if this construction, based on the `GRVV algebra',\footnote{The same defining matrix equation for the fuzzy sphere   appears while looking for BPS/ground state solutions in the pure/mass-deformed ABJM   theory. We will refer to it as the GRVV algebra throughout the rest of this paper. This equation first appeared as a BPS condition in \cite{Terashima:2008sy}, while its relation to the M2-M5 system was also investigated in \cite{Hanaki:2008cu}.} is indeed equivalent to the usual one in terms of the $\SU(2)$ algebra. We will find that there is indeed a one-to-one correspondence between the representations of the $\SU(2)$ algebra $J_i$ and the representations in terms of bifundamental matrices $\tilde G^\alpha$ satisfying the matrix equation of \cite{Gomis:2008vc}.  We will then relate this statement to the fact that the fuzzy supersphere of \cite{Grosse} is equivalent to the usual bosonic fuzzy sphere.  In \cite{Nastase:2009ny}, the relation between the matrices satisfying the GRVV algebra, and the matrix coordinates $J_i$ satisfying the $\SU(2)$ algebra, was the quantum (discretised) equivalent of the first Hopf map. Starting from this point we propose that the objects $\tilde G^\a$ defined on the fuzzy 2-sphere should be thought of as {\it fuzzy} versions of the Killing spinors on $S^2$.

Another issue that we wish to explore is the realisation of (twisted) supersymmetry in the context of the (fuzzy) 2-sphere as a solution to ABJM. In \cite{Nastase:2009ny} we obtained the large-$N$ action for small bosonic fluctuations on the 2-sphere. Here we complete the calculation by presenting the fermionic part of the fluctuation action, while obtaining some interesting subtleties. For the bosonic action, twisting the fields on the 2-sphere was a choice, and it was argued that the bosonic scalars transverse to the sphere should be related to 2-sphere twisted-spinors. However, the issue of twisting is tied in with the issue of supersymmetry. In this work we find that if one does not twist the fields on the sphere the action possesses a peculiar kind of $x$-dependence and supersymmetry, but if one twists the fields the $x$-dependence and supersymmetry are easy to understand. An interesting difference related to supersymmetry, that was not evident while studying the bosonic part of the fluctuation action, emerges between the fermionic pieces of the fuzzy sphere and the fuzzy funnel configurations. We also contrast our results with the previously analysed case of \cite{Andrews:2006aw}, for the `deconstruction' of a Maldacena-N\'u\~nez-type twisted compactification on $S^2$ from the Polchinski-Strassler ($\cN = 1^*$) 3+1d gauge theory \cite{Polchinski:2000uf}.

The rest of this paper is organised as follows. In Section\;\ref{review}, we review the fuzzy $S^2$ construction in 
terms of bifundamental matrices, as described in \cite{Nastase:2009ny}. In Section\;\ref{equivalence}, we prove the 
equivalence of this and the adjoint constructions in terms of the $\SU(2)$ algebra, and discuss its implications for 
the fuzzy supersphere.  In Section\;\ref{killing}, we identify the  GRVV matrices as fuzzy versions of Killing spinors 
on $S^2$.  In Section\;\ref{susyD4}, we proceed to find the supersymmetric D4-brane action for small fluctuations, both 
without and with a twisting of the fields on the sphere. We then compare with the deconstruction of the 
Maldacena-N\'u\~nez-type twisted compactification on $S^2$, as well as with the fuzzy funnel configuration.  In 
Section\;\ref{discussion} we conclude with a discussion, while in the Appendices we provide useful identities and 
conventions.

\section{Review of fuzzy $S^2$ construction from ABJM}\label{review}

We start by reviewing the fuzzy $S^2$ construction obtained in \cite{Nastase:2009ny} by studying  the  ground state solution of the maximally supersymmetric massive deformation of the ABJM model as given in \cite{Gomis:2008vc}.  Similarly, one can obtain the BPS fuzzy funnel for the pure ABJM model in terms of the above solution, through the replacement of the mass deformation parameter $\mu$ with $\frac{1}{2s}$, where $s$ is an M2 worldvolume coordinate along which the funnel grows into the M5. Although we have so far found no differences between the fuzzy sphere and fuzzy funnel solutions in the bosonic sector, we will see in Section\;\ref{susyD4} that the actions for fluctuations diverge in their fermionic part. This is as expected, since the funnel preserves half the amount of supersymmetry compared to the sphere.

Looking for ground-state/$\sfrac{1}{2}$-BPS solutions in the mass-deformed/pure ABJM theory leads to a simple set of solutions with $R^\a=f G^\a$, $Q^{\dot\a}=0$, where $C^I=(R^\a,Q^{\dot \a})$ form the 4 complex scalars of the ABJM model. Here $f=\frac{k\mu}{2\pi}$ for the sphere and $f=\frac{k}{4\pi s}$ for the 2-funnel. With the replacement $R^\a= f G^\a$, both the fuzzy sphere solution of the mass deformed theory and the fuzzy funnel solution of pure ABJM give the same equation for $G^\a$. We will take this as the defining equation (`algebra') for this bifundamental fuzzy $S^2$ construction, analogous to the $\SU(2)$ algebra relations for the usual fuzzy 2-sphere in terms of adjoint matrices.

This BPS/ground state matrix equation for $G^\a$ is
\be
-G^\a=G^\b G^\dagger_\b G^\a-G^\a G^\dagger_\b G^\b\;.\label{GRVV}
\ee
It is solved by the irreducible matrix representation $G^\a=\tilde{G}^\a$ of \cite{Gomis:2008vc}, 
\bea\label{BPSmatrices}
&& ( \tilde{G}^1)_{m,n }    = \sqrt { m- 1 } ~\delta_{m,n} \cr
&& ( \tilde{G}^2)_{m,n} = \sqrt { ( N-m ) } ~\delta_{ m+1 , n } \cr
&& (\tilde{G}_1^{\dagger} )_{m,n} = \sqrt { m-1} ~\delta_{m,n} \cr
&& ( \tilde{G}_2^{\dagger} )_{m,n} = \sqrt { (N-n ) } ~\delta_{ n+1 , m }\;.
\eea
Since the $G^\alpha$'s transform in the bifundamental of the $\U(N)\times\U(\bar N)$ gauge group of  ABJM, they are $N\times \bar N$ matrices. Equivalently, the $G^\dagger_\alpha$'s are $\bar N \times N$ matrices.  Defining the $\U(2)$ symmetry generators ${J^\a}_\b=G^\a G^\dagger_\b$ and ${\bar J_\a\,}^\b=G^\dagger_\a G^\b$, one can extract the usual $\SU(2)$ components by considering
\bea\label{defining}
J_i&=&{(\tilde\s_i)^\a}_\b G^\beta G^\dagger_\alpha={(\tilde\s_i)^\a}_\b{J^\b}_\a\equiv{(\s_i)_\b}^\a {J^\b}_\a\;\cr
\bar{J}_i&=&{(\tilde\s_i)^\a}_\b G^\dagger_\alpha G^\beta =  {(\tilde{\s}_i)^\a}_\b {\bar{J}_\a\,}^\b\equiv{(\sigma_i)_\b}^\a{\bar{J}_\a\,}^\b\;.
\eea
We note that the $J_i$ act on an $N$-dimensional vector space, which is an irreducible representation of $\SU(2)$ and we call ${\bf V}^+$, while $\bar J_i$ act on an $N$-dimensional vector space ${\bf V}^-=V^-_{N-1}\oplus V_1^-$, which is a reducible representation of $\SU(2)$, as the sum of an $(N-1)$- and a 1-dimensional representation with an element $E_{11}$, acting on $|e_1^-\rangle$  \cite{Nastase:2009ny}.

One easily finds that the $G^\a$, as well as all bifundamental fields, transform under the combined action
\be
J_i G^\a -G^\a \bar J_i={(\ts_i)^\a}_\b G^\b\;.\label{mixed}
\ee
As a result, a \emph{single}, diagonal $\SU(2)$ subgroup survives as a symmetry of the system.

In the classical limit ($N\to \infty $), $x_i=\frac{J_i}{\sqrt{N^2-1}}$ and $\bar x_i=\frac{\bar J_i}{\sqrt{(N-1)^2-1}}$ play the role of the same Euclidean coordinate on the 2-sphere. 
Then the defining relation (\ref{defining}) becomes $x_i = \bar x_i =g^\dagger_\a \:{(\ts_i)^\a}_\b\: g^\b $. 
If $g^\a$ are classical limits of general solutions of (\ref{GRVV}), satisfying only $g^\a g^\dagger_\a=1$, then 
the relation is 
the usual first Hopf map $S^3\stackrel{\pi}{\rightarrow} S^2$, from the 3-sphere $g^\a g^\dagger_\a =1$ onto the 
2-sphere $x_i x_i=1$. However, since in (\ref{BPSmatrices}) one has $\tilde{G}^1 = \tilde{G}^\dagger_1$, 
the irreducible GRVV matrices actually encode three 
real degrees of freedom, appropriate for an $S^2$, as opposed to the four needed for an $S^3$. This is in agreement 
with the expectation from the $\SU(2)$ symmetry structure. We will revisit this statement in greater detail in 
Section\;\ref{equivalence}. 

The $S^2$ picture can also be verified by a small fluctuation analysis around the fuzzy sphere vacuum. In the classical limit, $J_i$ and $\bar J_i$ give not only symmetry operators, but also classical coordinates, while their adjoint action acts like a derivation on the (fuzzy) 2-sphere
\be
[J_i,.]\to-2i\epsilon_{ijk}x_j\d_k=-2i K_i^a\d_a\;,
\ee
with $K_i^a$  a set of Killing vectors on $S^2$, the precise definitions for which can be found in Appendix\;\ref{AppA}.

The scalar matrix fluctuations on the fuzzy 2-sphere, $r^\a=R^\a-fG^\a$, decompose  as
\bea
r^\a&=&r G^\a+{s^\a}_\b G^\b+T^\a\cr
{s^\a}_\b&=&\sfrac{1}{2}s_i{(\ts_i)^\a}_\b\;,
\eea
where $T^a$ have only nonzero elements $(T^\a)_{iN},(T^\a)_{Nj}$.
Using the standard map between matrix-valued fields on the fuzzy sphere and functions on $S^2$, we obtain in the classical limit
\bea
s_i&=&K_i^a A_a +x_i\phi;\;\;\; T^\a\rightarrow 0\cr
r^\a&=&K_i^a A_a\frac{{(\ts_i)^\a}_\b}{2}G^\b+\frac{(2r+\phi)}{2}G^\a\;,
\eea
where now $A_a$ is a gauge field on $S^2$. We note that one scalar degree of freedom, $(2r-\phi)$ does not appear in the final action, as it has been `eaten up' by the 2+1d gauge fields through a large-$N$ version of the Higgs mechanism present for CS-matter theories \cite{Mukhi:2008ux}.\footnote{See also \cite{Nicolai:2003bp} for earlier work.} The Higgsing procedure, which renders the diagonal subgroup of the two Chern-Simons gauge fields $A^{(i)}_\mu$ dynamical, starts with the 
redefinition 
\bea\label{lincomb}
A_\mu &=&\sfrac{1}{2}({ A}^{(1)}_\mu +  { A}^{(2)}_\mu) \cr 
B_ \mu&=& \sfrac{1}{2}( { A}^{(1)}_\mu -  {  A}^{(2)}_\mu)\;,
\eea
after which $A_\mu$ becomes a $\U(1)$ Maxwell field on the 2-sphere, while $B_\mu$ is auxiliary and can be integrated out. The final action for the bosonic fluctuations once again reveals the $S^2$ structure, in terms of the bosonic part of an abelian 4+1d YM theory wrapped on the sphere \cite{Nastase:2009ny}.

\section{Equivalence of fuzzy sphere constructions and relation to fuzzy supersphere}\label{equivalence}

We now proceed to prove that the definition of the fuzzy 2-sphere in terms of bifundamentals is equivalent to the usual definition in terms of adjoint representations of the $\SU(2)$ algebra and that it implies the triviality of the fuzzy supersphere, in a way that we explain.

The matrix equation of motion for the fuzzy sphere background (\ref{GRVV}), 
can be rewritten using the U(2) generators as 
\be\label{meom}
-G^\a=G^\b {\bar J_\b\,}^\a-{J^\a}_\b G^\b
\ee
and is invariant under an $\U(N)\times \U(\bar N)$ gauge symmetry. This symmetry was used by \cite{Gomis:2008vc} to fix the irreducible $\tilde{G}^\a$ matrices that solve the above $G^\a$ equation
to the form given in the previous section, which in particular has 
$\tilde{G}^1=\tilde{G}^\dagger_1$. 

In the ABJM Lagrangean the bifundamental scalars were interpreted as Matrix Theory versions of Euclidean coordinates. Similarly, in the large $N$-limit one can write the matrices $G^\a\rightarrow \sqrt{N}g^\a$, with $g^\a$ for the moment 
as some commuting classical objects, to be identified and better understood in due course. 
In that limit the coordinates
\bea
x_i&=&{(\ts_i)^\a}_\b g^\b g^\dagger_\a\nonumber\\
\bar x_i&=& {(\ts_i)^\a}_\b g^\dagger_\a g^\b\label{hopf}
\eea
are two versions of the same Euclidean coordinate on the 2-sphere, $x_i\simeq \bar x_i$. 

Note that in the above construction the 2-sphere coordinates $x_i, \bar x_i$ in Eq.\;(\ref{hopf}) are invariant under multiplication of the classical objects $g^\a$ by a $\U(1)$ phase, thus we can define objects $\tilde{g}^\a$ \emph{modulo} such a phase, \ie $g^\a=e^{i\a(\vec{x})}\tilde g^\a$.  The GRVV matrices $\tilde{G}^\a$ are fuzzy versions of representatives of $\tilde g^\a$, chosen such that $\tilde{g}^1=\tilde{g}_1^\dagger$ (one could of course have chosen a different representative for $\tilde{g}^\a$ such that $\tilde{g}^2=\tilde{g}_2^\dagger$ instead).

It is in terms of the $ g^\alpha$'s that one has the usual Hopf map structure from the 3-sphere $ g^\a g^\dagger_\a=1$ onto the 2-sphere $x_ix_i=1$. In this picture, the phase is simply the coordinate on the $\U(1)$ fibre of the Hopf fibration, while the $\tilde g^\alpha$'s are coordinates on the $S^2$ base. While $g^\a$ are complex coordinates acted upon by $\SU(2)$, the $\tilde g^\a$ are real objects acted upon by the spinor representation of $\SO(2)$, so they can be thought of as Lorentz spinors in two dimensions, \ie spinors on the 2-sphere. This will become very important in Section\;\ref{killing}.

The fuzzy version of the full Hopf map, $J_i={(\ts_i)^\a}_\b G^\b G^\dagger_\a$, can be given either using $G^\a= U \tilde G^\a $ or $G^\alpha = \tilde{\hat   G}^\alpha\hat{U}$. The $U$ and $\hat{U}$ are unitary matrices that can themselves be expanded in terms of fuzzy spherical harmonics \be U=\sum_{lm} U_{lm} Y_{lm}(J_i)\;, \ee with $U U^\dagger = \hat U \hat U^\dagger =1$, implying that in the large-$N$ limit $(U,\hat U)\rightarrow e^{i\a(\vec{x})}$.

That means that by extracting a unitary matrix from the left or the right of $G^\a$, \ie modulo a unitary matrix, the resulting algebra for $\tilde{G}^\a$,
\be
-\tilde{G}^\a=\tilde{G}^\b \tilde{G}^\dagger_\b \tilde{G}^\a-\tilde{G}^\a \tilde{G}^\dagger_\b \tilde{G}^\b,
\label{algebra}
\ee
that we will call {\bf the GRVV algebra},
should then be exactly equivalent to the usual $\SU(2)$ algebra that appears in the adjoint construction: both should give the same description of the fuzzy 2-sphere. We would next like to prove this equivalence for all possible representations.

\subsection{Representations}

We first note that the irreducible representations of the algebra (\ref{algebra}), given in \cite{Gomis:2008vc} by the matrices (\ref{BPSmatrices}), indeed give the most general irreducible representations of SU(2). Defining $J_{\pm}=J_1\pm iJ_2$, $\bar J_{\pm}=J_1\pm i \bar J_2$, we obtain from (\ref{BPSmatrices}) that
\bea
(J_+)_{m,m-1}&=&2\sqrt{(m-1)(N-m+1)}=2\alpha_{\frac{N-1}{2},m-\frac{N+1}{2}}\nonumber\\
(J_-)_{n-1,n}&=&2\sqrt{(n-1)(N-n+1)}=2\alpha_{\frac{N-1}{2},n-\frac{N+1}{2}}\nonumber\\
(J_3)_{mn}&=&2\Big(m-\frac{N+1}{2}\Big)\delta_{mn}
\eea
and 
\bea
(\bar J_+)_{m,m-1}&=&2\sqrt{(m-2)(N-m+1)}=2\alpha_{\frac{N-2}{2},m-\frac{N+2}{2}}\nonumber\\
(\bar J_-)_{n-1,n}&=&2\sqrt{(n-2)(N-n+1)}=2\alpha_{\frac{N-2}{2},n-\frac{N+2}{2}}\nonumber\\
(\bar J_3)_{mn}&=&2\Big(m-\frac{N+2}{2}\Big)\delta_{mn}+N\delta_{m1}\delta_{n1}\;,
\eea
whereas the general spin-$j$ representation of $\SU(2)$ is
\bea
&&(J_+)_{m,m-1}=\alpha_{j,m}\nonumber\\
&&(J_-)_{n-1,n}=\alpha_{j,n}\nonumber\\
&&(J_3)_{mn}=m\delta_{mn}
\eea
(and the rest zero), where
\be
\alpha_{jm}\equiv \sqrt{(j+m)(j-m+1)}
\ee
and $m\in -j,...,+j$ takes $2j+1$ values. We note that the representation for $J_i$ is indeed the most general 
$N=2j+1$ dimensional representation, and since $(\bar J_+)_{11}=(\bar J_-)_{11}=(\bar J_3)_{11}=0$, the representation
for $\bar J_i$ is also the most general $(N-1)=2(j-\frac{1}{2})+1$ dimensional representation. 

However, we additionally have the $\U(1)$ generators completing the  $\U(2)$ symmetry, 
which in the case of the irreducible GRVV matrices $\tilde G^\a$ 
are diagonal and give the fuzzy sphere constraint $\tilde G^\a\tilde G^\dagger_\a\propto \one$, $\tilde G^\dagger _\a \tilde G^\a\propto \one$, 
\bea
&&J={J^1}_1+{J^2}_2=(N-1)\delta_{mn}\cr
&&\bar J ={\bar J_1\,}^1+{\bar J_2\,}^2=N\delta_{mn}-N\delta_{m1}\delta_{n1}\;,
\eea
where again $(\bar J)_{11}=0$, since $\bar J_i$ is in a $N-1\times N-1$ dimensional representation:
The element $E_{11}=\delta_{m1}\delta_{n1}$ is a special operator, so the first element of the vector space on which 
it acts is also special, \ie ${\bf V}^-=V^-_{N-1}\oplus V^-_1$. 

For a reducible representation of $\SU(2)$, the Casimir operator $\vec{J}^2=J_iJ_i$ giving the fuzzy sphere constraint is diagonal, with blocks proportional to the identity. The analogous object that  gives the fuzzy sphere constraint in our construction is the operator $J=G^\a G^\dagger_\a$. Indeed, in the case of reducible matrices modulo unitary transformations, $\tilde{G}^\a$, we find (in the same way as for $\vec{J}^2=J_i J_i$ for the $\SU(2)$ algebra)
\be
J=\text{diag}((N_1-1)\one_{N_1 \times N_1},(N_2-1)\one_{N_2 \times N_2},....)\label{j}
\ee
and similarly for $\bar J=G^\dagger_\a G^\a$
\be
\bar J=\text{diag}(N_1(1-E^{(1)}_{11})\one_{N_1 \times N_1}, N_2(1-E^{(2)}_{11})\one_{N \times N},...)\;.
\label{jbar}
\ee

\subsection{GRVV algebra $\rightarrow \SU(2)$ algebra}

For this direction of the implementation one does not need to consider the possible representations of the algebra; the matrices $\tilde{G}^\a$ will be kept as arbitrary solutions. We define as before, but now for an arbitrary solution $G^\a$,
\be
G^\a G^\dagger_\b\equiv {J^\a}_\b\equiv\frac{J_i {(\ts_i)^\a}_\b +J\delta^\a_\b }{2}\label{definiti}\;.
\ee
We additionally impose that $G^\a G^\dagger_\a\equiv J$ commutes with $J_k$.

Multiplying (\ref{GRVV}) from the right by ${(\ts_k)^\gamma}_\a G^\dagger_\gamma$, one obtains 
\be
-J_k = G^\b G^\dagger _\b J_k-{J^\a}_\b {J^\b}_\gamma {(\ts_k)^\gamma}_\a\;.
\ee
Using the definition in (\ref{definiti}) for the ${J^\a}_\b$ factors and the condition $[J,J_k]=0$, one
arrives at
\be
-J_k=\frac{i}{2}\epsilon_{ijk} J_j J_k\;,
\ee
which is just the SU(2) algebra. 

It is also possible to define
\be G^\dagger_\a G^\b\equiv {\bar J_\a\,}^\b\equiv \frac{\bar   J_i{(\ts_i)^\b}_\a+\bar J\delta^\b_\a}{2}
 \ee
 and impose the condition $[\bar J,\bar J_k]=0$. By multiplying (\ref{GRVV}) from the left by ${(\ts_k)^\gamma}_\a G^\dagger_\gamma$, we get in a similar way \be -\bar J_k=\frac{i}{2}\epsilon_{ijk}\bar J_i \bar J_k\;.  \ee

Thus the general SU(2) algebras for $J_i$ and $\bar J_i$ indeed follow immediately from (\ref{GRVV}) without restricting to the irreducible GRVV matrices.

\subsection{$\SU(2)$ algebra $\rightarrow$ GRVV algebra}

This direction of the implementation is {\it a priori} more problematic since, as we have already seen, the representations of $J_i$ and $\bar J_i$ are not independent. For the irreducible case in particular,  $V_N^+$ is replaced by the representation $V^-_{N-1}\oplus V^-_1$, so now we need to define this identification in the general case.

We will first try to understand 
the classical limit. The Hopf fibration (\ref{hopf}) can be rewritten, together with the normalisation 
condition, as
\be
 g^\alpha g^*_\beta = \frac{1}{2} \Big[x_i {(\ts_i)^\alpha}_\beta + \delta^\alpha_\beta\Big]  \;.
\ee

By extracting a phase out of $g^\a$, we should obtain the variables $\tilde{g}^\a$ on $S^2$ instead of $S^3$. 
Indeed, the above equations can be solved  for $g^\a$ by
\be
g^\alpha = \doublet{g^1}{g^2} = \frac{e^{i \phi}}{\sqrt{2(1+x_3)}}{\doublet{1+x_3}{x_1-ix_2}}=e^{i\phi}\tilde{g}^\a\;,
\label{tildeg}
\ee
where $e^{i\phi}$ is an arbitrary phase.

In the fuzzy case $G^\a$ and $G^\dagger_\b$ do not commute, and there are two different kinds of equations corresponding to $J_i$ and $\bar J_i$,
\bea
G^\alpha G^\dagger_\beta &\equiv& \frac{1}{2} \Big[J_i {(\ts_i)^\alpha}_\beta + \delta^\alpha_\beta J\Big]  \cr
G^\dagger_\beta  G^\alpha &\equiv& \frac{1}{2} \Big[\bar J_i {(\ts_i)^\alpha}_\beta + \delta^\alpha_\beta \bar J\Big]\;.
\label{reverse}  
\eea
We also impose as before that $[J,J_k]=0$, $[\bar J,\bar J_k]=0$, so that $J$ and $\bar J$ are diagonal and proportional to the identity in the irreducible components of $J_i$.

One solves the first set of equations in (\ref{reverse}) by  considering $G^1G^\dagger_1=\frac{1}{2}(J+J_3)$, for which the 
most general solution is $G_1 = T U$, with $T$  a Hermitian and $U$ a unitary matrix.
Since $J+J_3$ is a real and diagonal, by defining 
\be
T = \frac{1}{\sqrt 2}\Big(J+ J_3\Big)^{1/2}\;,
\ee
one  obtains
\be
G^\alpha = \doublet{G^1}{G^2} = {\doublet{J+J_3}{J_1-iJ_2}}\frac{T^{-1}}{2}U_{N\times\bar N}
=\tilde{G}^\alpha U_{N\times N}\;.
\ee
Thus $\tilde{G}^\alpha$ is also completely determined by $J_i, J$. 

Similarly,  the second set of equations in (\ref{reverse}) can be solved by  considering $G^\dagger_1 G^1=\frac{1}{2} (\bar J+\bar J_3)$, for which the most general solution is $G^1=\hat{U}\tilde{T}$, where as before
\be
\tilde{T} = \frac{1}{\sqrt 2}\Big(\bar J+ \bar J_3\Big)^{1/2}\;,
\ee
to obtain 
\be
G^\alpha = \doublet{G^1}{G^2} = \hat{U}_{N\times\bar N}
\frac{\tilde{T}^{-1}}{2}{\doublet{\bar J+\bar J_3}{\bar J_1-i\bar J_2}}=\hat{U}\tilde{\hat G}^\a\;.
\ee
Thus $\tilde{\hat G}^\a$ is completely determined by $\bar J_i, \bar J$.

Comparing the two formulae for $G^\a$ we see that they are compatible if and only if
\be
\hat{U}=TU\tilde{T}^{-1}\qquad\text{and}\qquad
\bar J_1-i\bar J_2=\tilde{T}^2U^{-1}T^{-1}(J_1-iJ_2)T^{-1}U\label{equiva}\;,
\ee
where $U$ is an arbitrary unitary matrix. These equations define an identification between the two representations of $\SU(2)$, in terms of $J_i$ and $\bar J_i$, needed  in order to establish the equivalence with the GRVV matrices.

In terms of explicit representations: for the irreducible representations of $\SU(2)$, we define $\bar J_i$ from $J_i$ as before ($V_N^+\rightarrow V_{N-1}^-\oplus V_1^-$) and $J=(N-1)\one_{N \times N}$, $\bar J=N(1-E_{11})\one_{N   \times N}$.  For reducible representations of $\SU(2)$, $J_i$ can be split such that $J_3$ is block-diagonal, with various irreps added on the diagonal.  One must then take $J$ and $\bar J$ of the form in (\ref{j}) and (\ref{jbar}).

Then the condition (\ref{equiva}) is solved by $U=1$ and $J_1,J_2$ block diagonal, with the blocks being the irreps of dimensions $N_1,N_2,N_3,...$, and the $\bar J_1, \bar J_2$ being also block diagonal, but where each $N_k\times N_k$ irrep block is replaced with the $(N_k-1)\times (N_k-1)$ irrep block, plus an $E^{(k)}_{11}$, just as for the GRVV matrices.

\subsection{Fuzzy superalgebra}
It is easy to see that the matrices $\tilde G^\a$ and $J_i$ can be neatly  packaged into supermatrices which form a representation of the orthosymplectic Lie superalgebra $\text{OSp}(1|2)$, and thus form supersymmetric partners. The supermatrix is nothing but the embedding of of the $N\times \bar N$ matrices into $\U(2N)$. The adjoint fields live in the  `even subspace', while the bifundamentals in the `odd subspace'.  For a generic supermatrix
\be
M = \twobytwo{A}{B}{C}{D}
\ee
the superadjoint operation is 
\be
M^\ddagger = \twobytwo{A^\dagger}{C^\dagger}{-B^\dagger}{D^\dagger}
\ee
For Hermitian supermatrices this is
\be
X = \twobytwo{A}{B}{-B^\dagger}{D}\;,
\ee
with $A = A^\dagger$ and $D = D^\dagger$ \cite{Hasebe:2004yp}. This gives the definition of the supermatrices
\be
{\bf J}_i = \twobytwo{J_i}{0}{0}{\bar J_i}\qquad\textrm{and}\qquad {\bf J}_\alpha = \twobytwo{0}{\sqrt N \tilde G_\alpha}{-\sqrt N \tilde G^\dagger_\alpha}{0}\;,
\ee
where we raise and lower indices as $\tilde G_\alpha = \epsilon_{\alpha\beta} \tilde G^\beta$, 
with $\epsilon = i \ts_2 = -i\s_2$. Then the SU(2) algebra together with the relation (\ref{mixed}) 
and the definition of $J_i, \bar J_i$ result in the following (anti)commutation relations
\bea\label{superalgebra}
[{\bf J}_i,{\bf J}_j] &=& 2 i \epsilon_{ijk} {\bf J}_k\cr
[{\bf J}_i,{\bf J}_\alpha] &=& {(\ts_i)_\alpha}_\beta {\bf J}^\beta\cr
\{{\bf J_\alpha},{\bf J_\beta}\} &=& - (\tilde \sigma_i)_{\alpha\beta} {\bf J}_i= - (i\tilde \sigma_2 \tilde \sigma_i)_{\alpha\beta} {\bf J}_i\;,
\eea
which is the defining superalgebra $\text{OSp}(1|2)$ for the fuzzy {\it supersphere} of \cite{Grosse}.\footnote{This observation has also been made in \cite{Maldacena:2009mw}.}

The emergence of the fuzzy supersphere might be a bit of a surprise here, since we have just shown that the GRVV and adjoint matrix constructions are actually equivalent. On the other hand, it is known that the only irreducible representations of $\text{OSp}(1|2)$ split into the spin-$j$ plus the spin-$(j-\frac{1}{2})$ representations of $\SU(2)$, which correspond {\em precisely} to the irreducible representation for the $\tilde G^\a$ that we are considering here.\footnote{See for instance Appendix C of \cite{Hasebe:2004yp}. The general spin-$j$ is the $J_i$ representation constructed from the GRVV matrices, while the general spin $j-\frac{1}{2}$ is the $\bar J_i$ representation constructed from the GRVV matrices.}

As a result, the most general representations of the fuzzy superalgebra coincide with the most general representations of the $\tilde G^\a$ themselves,  which as we showed are completely equivalent to the representations of $\SU(2)$. In other words, the statement is that the fuzzy supersphere is trivial, and contains the same information as the bosonic fuzzy sphere.

\subsection{$N\times M$ representations and the ABJ model}

An interesting related question is whether one gains anything qualitatively new by going to the $\U(N)\times \U(M)$ CS-matter theories of the Aharony-Bergman-Jafferis (ABJ) model \cite{Aharony:2008gk}.\footnote{These gauge theories were initially considered in \cite{Hosomichi:2008jb}.} This is a natural extension to consider since the BPS/ground state matrix equation for the $N\times M$ matrices is again given by (\ref{GRVV}). 

By defining the $N\times N$ matrix ${J^\a}_\b=\tilde G^\a \tilde G^\dagger_\b$ and the $M \times M$ matrix ${\bar J_\a\,}^\b=\tilde G^\dagger_\a \tilde G^\b$, with $J_i={J^\a}_\b{(\ts_i)^\b}_\a$, $\bar J_i={\bar J_\a}^\b {(\ts_i)^\a}_\b$, one might think that we could have $J_i$ being an irreducible $N\times N$ representation and $\bar J_i$ an irreducible $M\times M$ representation of $\SU(2)$. However, that would in turn mean that there is  both an $N\times N$ matrix 
\be
{J^1}_1=\tilde G^1\tilde G^\dagger_1=(m-1)\delta_{mn}
\ee
 and an $M\times M$ matrix
\be
{\bar J_1\, }^1=\tilde G^\dagger_1 \tilde G^1=(m-1)\delta_{mn}\;.
\ee
If an $N\times M$ matrix $\tilde G^1$, with  elements 
$a_{mn}$, that satisfied both relations existed then
\be
\sum_{i=1}^N a_{mi}a_{ni}=(m-1)\delta_{mn}\qquad\text{and also}\qquad
\sum_{j=1}^M a_{jm}a_{jn}=(m-1)\delta_{mn}\;.
\ee
This would imply (if, say $N<M$) that there exist $M$ linearly independent vectors of $M>N$ components, which is not possible.

Another related observation is that if such a $\tilde G^\a$  exists, again for $N<M$, it would be possible to reduce $M\times M$ irreps in terms of $N\times M$ ones. It can indeed be checked that the maximal irreducible representation is 
\be
\tilde G^\a_{N\times M}=(g^\a_{N\times N}|0_{N\times (M-N)}),
\ee
\ie the usual $N\times N$ irrep. However, if $M=rN+p$ with $r,p$ integers, then the representation
\be
G^\a_{N\times M}=\frac{1}{\sqrt{r}}(g^\a_{(1) N\times N}|...|g^\a_{(r)N\times N}|0_{N\times p})\label{solut}
\ee
is also a solution, if $g_{(i)N\times N}$ is the $N\times N$ solution.
This gives
\bea
{J^\a}_\b&=&({j^\a}_\b)_{N\times N}\nonumber\\
{\bar J_\a\,}^\b&=& \begin{pmatrix} ({\bar j_\a\,}^\b)_{N\times N}& ... &({\bar j_\a\,}^\b)_{N\times N}
&0\\ \vdots&\vdots&\vdots&\vdots\\ 
({\bar j_\a\,}^\b)_{N\times N} & ...&({\bar j_\a\,}^\b)_{N\times N}&0\\0&...&0&0\end{pmatrix}
\eea
\ie the $J_i$ representation is the $N\times N$ representation and $\bar J_i$ representation is made of $r$ copies of the $(N-1)\times (N-1)$ representation embedded in $N\times N$, plus zeroes for the rest. This also is nothing but another kind of reducible representation that one could consider. Therefore, nothing new is obtained by considering $N\times M$ matrices and the ABJ model.

This is in agreement with expectations from the spacetime interpretation of fuzzy sphere solutions in  mass-deformed ABJ theories. In the undeformed ABJ case, one has (say for $M>N$) $N$ M2-branes probing the $\mathbb Z_k$ singularity of M-theory and  $|M-N|$ \emph{fractional} M2-branes, corresponding to M5-branes wrapping a collapsed $S^3/\mathbb Z_k$ \cite{Aharony:2008gk}. While the $N$ M2's are free to move, the $|M-N|$ fractional M2's are forced to remain at the orbifold fixed point. In the mass-deformed case this would mean that the $N$ moving M2's can puff up into a fuzzy sphere configuration with the remaining fractional M2's stuck at the origin. In the gauge theory this is reflected by the fact that one only has solutions by giving vevs at most to an $N\times N $ block inside $N\times M$. This is precisely what we have found above.

\section{Fuzzy Hopf fibration \& fuzzy Killing spinors}\label{killing}

In this section we want to interpret the classical objects $\tilde{g}^\a$, obtained in the large-$N$ limit of $\tilde{G}^\a$, as Killing spinors on the 2-sphere and generalise this construction to higher dimensional cases.

We have seen that the in the classical limit, the relation between $J_i$ and $G^\a$ becomes the first Hopf map (\ref{hopf}), and hence can be thought of as a {\it fuzzy} version of the same. However, the above Hopf relation is invariant under multiplication by an arbitrary phase corresponding to shifts on the $S^1$ fibre, so the objects $\tilde{g}^\a$ obtained by extracting that phase in (\ref{tildeg}), \ie  
\be
 \tilde{g}^\a=\frac{1}{\sqrt{2(1+x_3)}}{\doublet{1+x_3}{x_1-ix_2}}\;,
 \ee
 are instead defined on the classical $S^2$. In the Hopf fibration, the index of $g^\alpha$ is a spinor index of the global $\SO(3)$ symmetry for the 2-sphere. By extracting the $S^1$ phase one recovers the real $\tilde{g}^\a$ and the $\a$ can be thought of as describing a (Majorana) spinor of the $\SO(2)$ local Lorentz invariance on the 2-sphere. We will argue that the latter is related to a Killing spinor. Note that this type of identification easily extends to all even spheres.

In the fuzzy version of this relation, the $\tilde{G}^\a$ obtained from $G^\alpha$ by extracting a unitary matrix, are real objects 
defined on the fuzzy $S^2$ through the GRVV matrices, in the case of irreducible representations, or 
\be
\tilde{G}={\doublet{J+J_3}{J_1-iJ_2}}\frac{T^{-1}}{2}
\ee
in general.

The standard interpretation, inherited from the examples of the $\SU(2)$ fuzzy 2-sphere and other spaces, is that the matrix indices give rise to the dependence on the sphere coordinates and the index $\a$ is a global symmetry index. However,  we have just seen that already in the classical picture one can identify the global symmetry spinor index with the local Lorentz spinor index. Therefore we argue that the correct interpretation of the classical limit for $\tilde{G}^\a$ is as a spinor with both global and local Lorentz indices, \ie the Killing spinors on the sphere $\eta^{\a I}$. In the following but we will use the index $\a$ interchangeably for the two.

For comparison with the Killing spinors, we write for the classical limit of the $J_i$-$\tilde{G}^\a$ relation as
\be
x_i\simeq\bar x_i = {(\sigma_i)_\a}^\b \tilde{g}^\dagger_\b \tilde{g}^\a\label{classi}\;.
\ee

\subsection{Killing spinors on $S^n$ and the special case of $S^2$}

Let us review some of the key facts about Killing spinors that we will need for our discussion. For more details, we refer the interested reader to \eg \cite{vanNieu1983,Eastaugh1985,vanNieuwenhuizen:1984iz,Gunaydin:1984wc,Nastase:1999kf}.

On a general sphere $S^n$, one has Killing spinors satisfying
\be
D_\mu \eta(x)=\pm \frac{i}{2}m\gamma_\mu \eta(x)\;,
\ee
where by calculating $[D_\mu,D_\nu]$ we obtain the normalisation of the curvature as
\be
{R_{\mu\nu}}^{mn}=m^2(e_\mu^m e_\nu ^n-e_\mu^n e_\nu^m)\;.
\ee
There are two kinds of 
Killing spinors, $\eta^+$ and $\eta^-$, which in even dimensions are related by the chirality matrix, \ie $\gamma_{n+1}$, through $\eta^+=\gamma_{n+1} \eta^-$, as  can be easily checked. The charge conjugation matrix in $n$ dimensions satisfies in general
\be
C^T=\kappa C;\;\;\;\;
\gamma_\mu^T=\lambda C \gamma_\mu C^{-1}\;,
\ee
where $\kappa=\pm, \lambda=\pm$ and it is used to raise/lower indices. The Majorana condition is then given by
\be
\bar \eta =\eta^T C\;.
\ee

The Killing spinors on $S^n$ satisfy orthogonality, completeness and a reality condition. The latter depends on the application, sometimes taken to be the \emph{modified} Majorana condition, which mixes (or identifies) the local Lorentz spinor index with the global symmetry spinor index of $S^n$. 
For instance, on $S^4$ the orthogonality and completeness are respectively,
\be
\bar \eta^I \eta^J=\Omega^{IJ}\;\qquad\text{and}\qquad
\eta^\a_J\bar \eta^J_\b=-\delta_\b^\a\;,
\ee
where the index $I$ is an index in a spinorial representation of the $\SO(n+1)_G$ invariance group of the sphere and the index $\a$ is an index in a spinorial representation of the $\SO(n)_L$ local Lorentz group on the sphere. The indices are then identified by the \emph{modified} Majorana spinor condition as follows\footnote{For more details   on Majorana spinors and charge conjugation matrices see   \cite{vanNieu1983,VanNieuwenhuizen:1981ae} and the Appendix of \cite{Nastase:1999kf}.}
\be
\bar \eta^I\equiv (\eta^I)^TC^{(n)}_{-}=-(\eta^J)^\dagger \gamma_{n+1} \Omega^{IJ}\;,
\ee
where $\Omega^{IJ} = i \sigma_2 \otimes \one_{\sfrac{n}{2}}$ is the invariant tensor of $\text{Sp}(\sfrac{n}{2})$, satisfying $\Omega^{IJ}\Omega_{JK} = \delta^I_K$.

The Euclidean coordinates of $S^n$ are bilinear in the Killing spinors 
\be\label{bilinear}
x_i=(\Gamma_i)_{IJ}\bar \eta^I \gamma_{n+1} \eta^J\;,
\ee
where $\eta$ are of a single kind (+ or -), or equivalently $\bar \eta_+^I\eta_-^J$.
For even $n$-spheres, we obtain the special case that the two spinor indices $\a$ and $I$ are of the same type, and 
we can write
\be
\eta^{\a I}={\Big\{\exp\Big(-\frac{i}{2}x^\mu \delta^m_\mu \gamma^m\Big)\Big\}^\a}_\b\eta^{\b I}(0)\;,
\ee
where $\eta^{\b I}(0)=\epsilon^{\b I}$ a constant spinor. 

Starting from  Killing spinors on $S^n$, one can construct all the higher spherical harmonics. As seen in Eq.\;(\ref{bilinear}), Euclidean coordinates on the sphere are spinor bilinears. In turn, symmetric traceless products of the $x_i$'s  construct the scalar spherical harmonics $Y^k(x_i)$.\footnote{These are the higher dimensional extensions of the usual spherical harmonics $Y^{lm}(x_i)$ for $S^2$.}  One can also construct the set of spinorial spherical harmonics by acting with an appropriate operator on $Y^k\eta^I$
\bea
\Xi^{k,+}&=& [(k+n-1+iD\!\!\!\!/)Y^k]\eta_+\cr
\Xi^{k,-}&=& [(k+n-1+iD\!\!\!\!/)Y^k]\eta_-=[(k+1+iD\!\!\!\!/)Y^{k+1}]\eta_+\;.
\eea
Note that in the above the derivatives act only on the scalar harmonics $Y^k$.  

Any spinor on the sphere can be expanded in terms of spinorial spherical harmonics, $\Psi=\sum_k \psi_k\Xi^{k,\pm}$. Consistency imposes that the $\Xi^{k,\pm}$ can only be commuting spinors. The Killing spinors are then themselves {\em commuting} spinors, as they are used to construct the spinorial spherical harmonics. 

For higher harmonics the construction extends in a similar way but the formulae are more complicated and, as we will not need them for our discussion, we will not present them here. The interested reader can consult \eg \cite{Kim:1985ez}.

\subsubsection*{Killing spinors on $S^2$}

For the particular case of the $S^2$, $\gamma_i=\Gamma_i = \sigma_i$ for both the $\SO(2)_L$ and the $\SO(3)_G$ Clifford algebras. Then the two $C$-matrices can be chosen to be: $C_+=-\sigma_1$, giving $\kappa=\lambda=+$, and $C_-=i\sigma_2=\epsilon$, giving $\kappa=\lambda=-$. Note that with these conventions one has $C_-\gamma_3=i\sigma_2\sigma_3=-\sigma_1=C_+$. In the following we will choose the Majorana condition to be defined with respect to $C_-$.

 Eq.\;(\ref{bilinear}) then gives for $n=2$
\be
\bar \eta^I =(\eta^T)^I C_-\Rightarrow x_i=(\sigma_i)_{IJ}(\eta^T)^I C_+ \gamma_3 \eta^J\label{classiK}\;.
\ee
The orthonormality and completeness conditions for the Killing spinors on $S^2$ are 
\be
\bar \eta^I \eta^J=\epsilon^{IJ}\qquad\text{and}\qquad\eta^\a_J \bar \eta^J_\b =-\delta^\a_\b\;,
\ee
while the modified Majorana condition is 
\be
(\eta^J)^\dagger=\epsilon_{IJ}\bar \eta^I\equiv \epsilon_{IJ}(\eta^I)^T C_-\;.
\ee
Since $C_-=\epsilon$, by making both indices explicit and by renaming the index $I$ as $\dot\a$ for later use, one also  
has
\be\label{modifiedeta}
(\eta^{\a\dot\a})^\dagger=\eta_{\a\dot\a}\equiv \epsilon_{\a\b}\epsilon_{\dot\a\dot\b}\eta^{\b\dot\b}\;.
\ee

Using this condition, we rewrite (\ref{classiK}) as 
\be
x_i={(\sigma_i)^I}_J (\eta^I)^\dagger \gamma_3\eta^J
={(\ts_i)^I}_J\Big(\sqrt{2}P_+\eta^I\Big)^\dagger\Big(\sqrt{2}P_+\eta^J\Big)\;,
\ee
where $P_\pm = \frac{1}{2}(1\pm\gamma_3)$. Now comparing (\ref{classiK}) with (\ref{classi}) one is led to the following natural large-$N$ relation, $\tilde{G}^\a\rightarrow \sqrt{2N} P_+\eta^I$, provided the spinor indices $\a$ and $I$ get identified, \ie
\be\label{identification}
\frac{\tilde{G}^\a}{\sqrt{N}}\equiv \tilde{g}^\a\leftrightarrow \tilde{g}^I\equiv \sqrt{2}P_+\eta^I\;.
\ee
Note that the Weyl projection kills the omitted $\a$ spinor index on $\eta^I$. We will investigate the above expression more thoroughly in the next 
subsection, where we will also see that there is one more subtlety related to this identification.

Finally, the spinorial spherical harmonics on $S^2$ are 
\be
\Xi^{\pm}_{lm}=[(l+1+iD\!\!\!\!/\;)Y_{lm}]\eta_{\pm}\label{killspsphhar}
\ee
and thus the spherical harmonic expansion of an $S^2$-fermion is (writing explicitly the sphere fermionic index $\a$)
\be
\psi^\a=\sum_{lm,\pm} \psi_{lm,\pm}\Xi^{\pm,\a}_{lm}=\sum_{lm,\pm} {[\psi_{lm,\pm}(l+1+iD\!\!\!\!/\;)Y_{lm}]^\a
}_\b\eta_{\pm}^\b\;.
\ee

\subsection{Relation between spinors on the 2-sphere}

On the 2-sphere, one defines the Killing vectors $K_i^a$ such that the adjoint action of the $\SU(2)$ generators on the fuzzy sphere fields becomes a derivation in the large-$N$ limit\footnote{The precise expressions for the Killing vectors $K_i^a$ can be found in Appendix\;\ref{AppA}.}
\be
[J_i,.]\to 2iK_i^a\d_a=2i\epsilon_{ijk}x_j\d_k\;.
\ee
One can then explicitly check that 
\be
K_i^a{(\ts_i)^\a}_\b=-e^{am}{(S \sigma^m S^{-1})_\b\, }^\a\equiv -{(S \Gamma^a S^{-1})_\a\,}^\b\label{srel}\;,
\ee
where $e^{am}$ is the vielbein on the sphere and
\be\label{Smatrix}
S=S(\phi)S(\theta)=a\begin{pmatrix}
-\sin{\sfrac{\theta}{2}} \,e^{i\phi/2}& -i\cos{\sfrac{\theta}{2}} \,e^{i\phi/2}\\
\cos{\sfrac{\theta}{2}} \,e^{-i\phi/2}&-i\sin{\sfrac{\theta}{2}} \,e^{-i\phi/2}
\end{pmatrix}\;,
\ee
with $|a|^2=1$ is a unitary matrix. The matrices
\be
S(\phi)=a_1\begin{pmatrix} 0 & -ie^{i\phi/2}\\ e^{-i\phi/2} &0\end{pmatrix}\qquad\;,\qquad S(\theta)=a_2\begin{pmatrix} \cos{\sfrac{\theta}{2}} &-i\sin{\sfrac{\theta}{2}}\\ -i\sin{\sfrac{\theta}{2}} & \cos{\sfrac{\theta}{2}}
\end{pmatrix}\;,
\ee
with $a=a_1a_2, |a_1|^2=|a_2|^2=1$ are rotation matrices, since a Lorentz rotation on spinors acts by
\be
{\Lambda^\mu}_\nu \gamma^\nu=S\gamma^\mu S^{-1}\;.
\ee
One can also impose the following (symplectic) reality condition on $S$
\be\label{symplectic}
\epsilon_{\a\beta}{(S^{-1})^\b}_\gamma \epsilon^{\gamma\delta}={(S^T)_\a}^\delta=
{S^\delta}_\a\;,
\ee
which fixes $a=\sqrt{i}^*$ and ensures that 
\be
{(S \sigma_i S^{-1})_\a\,}^\b={(S\sigma_i S^{-1})^\b}_\a\;,
\ee
since by explicit evaluation one can show that ${(\sigma_i)^\a}_\b={(\sigma_i)_\b\,}^\a$. Then it is also possible to check  that 
\bea
{(S\Gamma_3 S^{-1})^\a}_\b &=&-x_i{(\ts_i)^\a}_\b\cr
{(S\Gamma_a S^{-1})^\a}_\b &=&-h_{ab }K_i^b{(\ts_i)^\a}_\b\label{gamma3}\;.
\eea

The matrix ${S^\alpha}_\beta$ can additionally be used to go between spherical and Euclidean spinors on $S^2$. Because of the reality condition (\ref{symplectic}), if one also has real spinors obeying
\be
(\chi_{\a\dot\a})^\dagger=\chi^{\a\dot\a}\equiv \epsilon^{\a\b}\epsilon^{\dot\a\dot\b}\chi_{\b\dot\b},
\ee
which was identified in (\ref{modifiedeta}) as the {\em modified} Majorana spinor condition,
it follows that the ${S^\alpha}_\beta$-rotated spinors are themselves real, namely they obey
\be
((\chi_{\dot\a}S)_\a)^\dagger=(S^{-1}\chi^{\dot\a})^\a\equiv -\epsilon^{\dot\a\dot\b}(S^{-1})^{\a\b}
\chi_{\b\dot\b}=\epsilon^{\dot\a\dot\b}\epsilon^{\a\b}(\chi_{\dot\b}S)_\b\;. \label{rotreal} 
\ee

Next define 
\be\label{killingdef}
\eta^{I\a}={(S^{-1})^\a}_\b \eta_0^{I\b}= \frac{1}{\sqrt{2}} {(S^{-1})^\a}_\b \epsilon^{\b I}=\frac{1}{\sqrt{2}}
{S^I}_J \epsilon^{\a J},
\ee
where in the last equality we used the  (symplectic) reality condition on $S$. From (\ref{rotreal}) it is clear  that the $\eta^{I\a}$  obey the \emph{modified} Majorana condition, as was also required for Killing spinors. It is then possible to use (\ref{gamma3}) to prove that 
\be
x_i=(\gamma_i)_{IJ}\bar \eta^I \gamma_3 \eta^J\;,
\ee
hence verifying that the  $\eta^{I\a}$ are indeed Killing spinors. One can also explicitly check that 
\be
D_a({(S^{-1})^\a}_\b \epsilon^{\b I})=+\frac{i}{2}{(\Gamma_a)^\alpha}_\beta {(S^{-1})^\beta}_\gamma \epsilon^{\gamma I}\;,
\ee
which in turn means that 
\be
\frac{1}{\sqrt{2}} {(S^{-1})^\a}_\b \epsilon^{\b I}=\eta_+^{\a I}\;.
\ee 

According to the relation (\ref{identification}), the object to be matched against $\tilde{g}^\a$ is 
then 
\be
\sqrt{2}P_+\eta^I={(P_+)^\a}_\b {(S^{-1})^\b}_\g \epsilon^{\g I}
={(P_+)^\a}_\b {S^I}_J \epsilon^{\b J}={S^I}_J{(P_-)^J}_K
\epsilon^{\a K}\label{identindic}\;.
\ee
Thus, the Weyl projection can be thought of as `removing' either $\a$ or $I$, since only one of the two spinor components is non-zero.

In order to further check this proposed identification at large-$N$ we now calculate 
\be
\d_a (\sqrt{2}P_+\eta^I)= -\frac{i}{2} {(S \Gamma_a S^{-1})^I}_J 
(\sqrt{2}P_+\eta^J)
+\tilde{T}_a (\sqrt{2}P_+\eta^I)\label{killisp}\;,
\ee
where $\tilde{T}_\theta=0$ and $\tilde{T}_\phi=\frac{i}{2}\cos\theta$ and 
\be
(\d_a S )S^{-1}=-\frac{i}{2}S\Gamma_aS^{-1}+S T_a S^{-1}
\ee
by explicitly evaluation, with $T_\theta=0$ and $T_\phi=-\frac{i}{2}\cos\theta\; \gamma_3$.
 
This needs to be compared with the analogous result given in Eq.\;(4.48) of \cite{Nastase:2009ny} from the classical limit of the adjoint action of $J_i$ on $\tilde{G}^\a$, \ie from $[J_i,\tilde{G}^\a]$, where it was found that 
\bea
\d_a \tilde{g}^\a&=&\frac{i}{2}\hat h_{ab}K_i^b{(\ts_i)^a}_\beta \tilde g^\beta\cr &=& -\frac{i}{2}{(S\Gamma_a S^{-1})^\a}_\b \tilde{g}^\b\label{classg}\;.
\eea
In the second line we made use of the identity (\ref{gamma3}). In \cite{Nastase:2009ny} it was also verified that the above could reproduce the correct answer for $\d_a x_i$, which can be rewritten as 
\be
\d_a x_i=-\frac{i}{2}\tilde{g}^\dagger_\a\Big[{(\ts_i)^\a}_\b {(S\Gamma_a S^{-1})^\b}_\g-{(S\Gamma_a S^{-1})^\a}_\b {(\ts_i)^\b}_\g\Big]
\tilde{g}^\g\;.
\ee

Note that even though there is a difference between (\ref{killisp}) and (\ref{classg}), given by the purely imaginary term $\tilde{T}_a$ that is proportional to the identity, the two answers for $\d_a x_i$ exactly agree, since in that case the extra contribution cancels.  This extra term is a reflection of a double ambiguity: On one hand there is the extra index $\a$ on $\eta^I$, which can be acted upon by matrices, even though it is Weyl-projected, in effect multiplying the Weyl-projected $\eta^I$ by a complex number; if the complex number is a phase, it will not change any expressions where the extra index is contracted, thus we have an ambiguity against multiplication by a phase.  On the other, $\tilde{g}^\a$ is just a representative of the reduction of $g^\a$ by an arbitrary phase, so it is itself only defined up to a phase.  The net effect is that the identification of the objects in (\ref{killisp}) and (\ref{classg}) is only up to a phase. Indeed, locally, near $\phi\simeq 0$, one could write
\be
\tilde{g}^\a e^{\frac{i}{2}\phi \cos\theta}\leftrightarrow \sqrt{2}P_+\eta^I
\ee
but it is not possible to get an explicit expression for the phase over the whole sphere.

\subsection{Generalisations} 

On a general $S^{2n}$ some elements of the above analysis carry through. That is because even though it is possible to write for every $S^{2n}$ 
\be
x_A=\bar\eta^I (\Gamma_A)_{IJ} \gamma_{2n+1}\eta^J\;,
\ee
where $\eta^I$ are the Killing spinors, one only has fuzzy versions of the  quaternionic and octonionic Hopf maps to match it against. We will next find and interpret the latter in terms of Killing spinors on the corresponding spheres. 

\subsubsection{$S^4$}

The second Hopf map, $S^7\stackrel{\pi}{\rightarrow} S^4$, is related to the quaternionic algebra. Expressing the $S^7$ in terms of complex coordinates $g^\a$, now with $\a=1,...,4$, one has the sphere constraint have $g^\a g^\dagger_\a=1$ ($g^\a g^\dagger_\a=1\Rightarrow x_A x_A=1$; $A = 1,...,5$). The map in this case is (see for instance 
\cite{Wu:1988py})
\be
x_A=g^\b {(\Gamma_A)^\a}_\b g^\dagger_\a,
\ee
with ${(\Gamma_A)^\a}_\b$ the $4\times 4$ $\SO(5)$ gamma matrices.\footnote{These are constructed as: $\sigma_1$ and $\sigma_3$ where $1$ is replaced by ${\one}_{2\times 2}$ and 
$\sigma_2$ where $i$ is replaced by $i\sigma_1, i\sigma_2,i\sigma_3$.} Here we have identified the spinor index $I$ of $\SO(5)$ with the Lorentz spinor index $\a$ of $\SO(4)$. 

The $g^\a$'s start off as complex coordinates, being acted upon by $\SU(4)$, but projecting down to the base of the Hopf fibration we can replace $g^\a$ in the above formula with real $\tilde{g}^\alpha$'s, instead acted upon by the spinorial representation of $\SO(4)$, \ie by spinors on the 4-sphere. This process is analogous to what we saw for the case of the 2-sphere. Once again, it is possible to identify $\tilde{g}^\a$ with the Killing spinors, this time on $S^4$.

This suggest that one should also be able to write a spinorial version of the fuzzy 4-sphere 
for some bifundamental matrices $\tilde{G}^\a$, satisfying 
\bea
J_A&=&\tilde{G}^\b {(\Gamma_A)^\a}_\b \tilde{G}^\dagger_\a\cr
\bar J_A&=& \tilde{G}^\dagger_\a {(\Gamma_A)^\a}_\b \tilde{G}^\b\;,
\eea
where $J_A,\bar J_A$ should also play the role of $\SO(5)$ generators, that is they should satisfy
\be
J_A\tilde G^\a- \tilde G^\a \bar J_A={(\Gamma_A)^\a}_\b\tilde G^\b\;.
\ee
This in turn implies the same GRVV algebra as for the $S^2$ case
\be
\tilde G^\a=\tilde G^\a\tilde G^\dagger_\b\tilde G^\b -\tilde G^\b\tilde G^\dagger_\b\tilde G^\a
\ee
but now with $\tilde G^\a$ being 4 complex matrices that describe a fuzzy 4-sphere. We leave the investigation of this interesting possibility for future work.

\subsubsection{$S^8$}
The third Hopf map, $S^{15}\stackrel{\pi}{\rightarrow} S^8$, is related to the octonionic algebra. The $S^{15}$ is expressed now
by the real objects $g^T_\a g^\a=1$, $\a=1,...,16$ that can be split into two groups ($1,...,8$ and $9,...,16$). The Hopf map is expressed  by \cite{Bernevig:2003yz} ($g^T_\a g_\a=1\Rightarrow x_Ax_A=1$)
\be
x_A=g^T_\a(\Gamma_A)^{\a\b}g_\b\;,
\ee
where $(\Gamma_A)^{\a\b}$ are the $\SO(9)$ gamma-matrices.\footnote{The gamma-matrices are constructed 
similarly to the $S^4$ case as follows: $\Gamma_i=\begin{pmatrix}0&\lambda_i\\-\lambda_i &0\end{pmatrix}$, $\Gamma_8= 
\begin{pmatrix} 0&\one_{8\times 8}\\ \one_{8\times 8}&0\end{pmatrix}$, $\Gamma_9=\begin{pmatrix}\one_{8\times 8}&0\\0&-\one_{8\times
8}\end{pmatrix}$, \ie from $\sigma_2$ with $\lambda_i$ replacing $i$, and from $\sigma_1$ and $\sigma_3$ with 
$1$ replaced by $\one_{8\times 8}$. The
$\lambda_i$ satisfy $\{\lambda_i,\lambda_i\}=-2\delta_{ij}$ (similarly to the $i\sigma_i$ 
in the case of $S^4$) and are constructed from the structure constants of the algebra of the octonions \cite{Bernevig:2003yz}. An explicit 
inversion of the Hopf map is given by $g_\a=[(1+x_9)/2]^{1/2}u_\a$ for $\a=1,...,8$ and $g_\a=[2(1+x_9)]^{-1/2}
(x_8-x_i\lambda_i)u_{\a-8}$ for $\a=9,..,16$, with $u_\a$ a real 8-component  $\SO(8)$ spinor satisfying $u^\a u_\a=1$
thus parametrising the $S^7$ fibre.}
Similarly for the case of the $S^4$ above, even though $g^\a$'s start off as being 16-dimensional 
variables acted by the spinor representation of $\SO(9)$, one can project down to the base of the Hopf fibration 
and replace the $g^\a$'s with real 8-dimensional 
objects on the 8-sphere $\tilde{g}^\a$.  Then the $\tilde{g}^\a$'s are identified with the Killing spinors of $S^8$.

This once again suggests that one should be able to write a spinorial version of the fuzzy 8-sphere for some 
bifundamental matrices $\tilde{G}^\a$ satisfying 
\bea
J_A&=&\tilde G_\a(\Gamma_A)^{\a\b}\tilde G^T_\b\cr
\bar J_A&=&\tilde G^T_\a(\Gamma_A)^{\a\b}\tilde G_\b\;,
\eea
where $J_A,\bar J_A$ are $\SO(9)$ generators 
\be
J_A\tilde G_\a - \tilde G_\a \bar J_A={(\Gamma_A)_\a}^\b\tilde G_\b
\ee
and  implies the same GRVV algebra, but with the $\tilde G^\a$'s now being 16 dimensional real matrices that 
describe the fuzzy 8-sphere.

\subsubsection{$\cp^3$}

The first Hopf map, $S^3\stackrel{\pi}{\rightarrow} S^2$, can also be generalised to the $S^7\stackrel{\pi}{\rightarrow} \cp^3$ case, and thus the extension to the fuzzy level would imply generalising the fuzzy Killing spinors on $S^2\simeq\cp^1$ to fuzzy Killing spinors on $\cp^3$. For that, we first notice that the $S^3\stackrel{\pi}{\rightarrow} S^2$ map is better understood as $S^3\stackrel{\pi}{\rightarrow} \cp^1$ \cite{Nakahara:1990th}. Indeed, the $S^3$ coordinates $Z^\a$ ($Z^1=X^1+iX^2,Z^2=X^3+iX^4$) obey $\sum_\a |Z^\a|^2=1$, a relation  invariant under multiplication by a phase, $Z^\a\rightarrow e^{i\varphi} Z^\a$, which is precisely the $\U(1)$ fibre of the Hopf fibration. This can be seen as follows: The stereographically projected coordinates on the $S^2$ are
 \be
W=\frac{x_1+ix_2}{1-x_3}=\frac{X^1+iX^2}{X^3+iX^4}=\frac{Z^1}{Z^2}\label{stereo}\;,
\ee
which are obviously invariant under $Z^\a\rightarrow e^{i\varphi}Z^\a$. But they are also invariant under the more 
general condition $Z^\a\rightarrow \lambda Z^\a$, with $\lambda$ an arbitrary complex number, which means that 
(\ref{stereo}) is really a map between $\cp^1$ and $S^2$. Thus for the Hopf map one 
really first maps the $S^3$ in Euclidean coordinates $Z^a$ to the $\cp^1$ with the same coordinates (now identified
with any complex $\lambda$), which is a linear relation, obtained just by an equivalence
\be
\{ Z^\a|\sum_\a|Z^\a|^2=1 \}\rightarrow \{ Z^\a\sim \lambda Z^\a|\lambda \in \mathbb C-\{0\}\}\;.
\ee
Then the quadratic relation (\ref{hopf}) (with $g^\a\rightarrow Z^\a$) or the rational
stereographic relation (\ref{stereo}) can be thought of as a map between $\cp^1$ and $S^2$, or between $\SU(2)$-invariant coordinates $Z^\a$ and $\SO(3)$-invariant coordinates $x_i$, by means of the matrices ${(\ts_i)^\a}_\b$.

This implies a natural embedding for the $S^7\stackrel{\pi}{\rightarrow}\cp^3$ Hopf map. Indeed, in the classical case $S^7$ is similarly defined by $Z^i$, with $i=1,..,4$ and $\sum_{i=1}^4 |Z^i|^2=1$, and then the $\cp^3\simeq S^7/\mathbb{Z}_k|_{k\to\infty}$ is just obtained by the identification $Z^i\sim \lambda Z^i$. The restriction to $S^3\subset S^7$ is obtained by $Z^i\rightarrow (Z^1,Z^2)=Z^\a$, and similarly for $\cp^1\subset \cp^3$.

One can then construct the $S^7$ as a Hopf fibration $S^1\hookrightarrow  S^7\stackrel{\pi}{\rightarrow} \cp^3$ in terms of a quadratic relation, by using the 
complex $4\times 4$ matrices $\Sigma_M$ that take us between Euclidean coordinates $Z^i$ on the $S^7$ and 
local (unconstrained) coordinates $X_M$ on $\cp^3$
\be
X_M={(\Sigma_M)^j}_iZ^iZ^\dagger_j\;.
\ee
The $(\Sigma_M)_{ij}$ are  Clebsch-Gordan coefficients for the ${\bf 4}\otimes {\bf\bar 4}\rightarrow {\bf 6_A}$ product of the 
spinor representations into the vector of $\SO(6)\simeq \SU(4)$, hence $M=1,...,6$.\footnote{It is easy to see that 
this defines a fibration: $X_M$ is invariant under multiplication of the $Z^i$ by a $\U(1)$ phase, corresponding to the fibre. The $X_M$ are thus $\SO(6)$-invariant coordinates on the base.}
Now, by replacing the 4 complex coordinates $Z^i$ with the $\SU(4)$-invariant 6d 
Killing spinors $\eta^i_\sigma$,\footnote{The $\SO(6)$ Lorentz invariance group, 
with fermionic index $\sigma$, is now the same as the global $\SU(4)$ invariance group.} one can construct $\SO(6)$-invariant bosonic coordinates $X_M$ as before by 
\be
X_M=(\Sigma_M)_{ij}\bar \eta^i\gamma_7 \eta^j\;.
\ee
In the above relation, the Killing spinors on $\cp^3$, together with its metric, are inherited from the definition of
$\cp^3$ as the $k\rightarrow\infty$ limit of the $S^7/\mathbb Z_k$ reduction. Note that by taking a different metric on 
$\cp^3$, different Killing spinors are obtained \cite{bar}.

The fuzzy version of this relation would naturally  be 
\be
J_M={(\Sigma_M)^j}_i\tilde{Z}^i\tilde{Z}^\dagger_j\label{fuzzycp3}\;,
\ee
with $\tilde{Z}^i$ 4 complex matrices giving a fuzzy version of $\cp^3$ that reduce to the $\tilde{G}^\a$ matrices, describing the fuzzy $S^2\simeq\cp^1$, for $i=\a$.

It is not clear how one would construct a fuzzy $\cp^3$ algebra, or if it could arise as a solution of ABJM,
but the relation (\ref{fuzzycp3}) also defines variables $X_M$ on $\cp^3$ that are natural from the ABJM 
point of view, and should be important in the $\text{AdS}_4\times \cp^3$ / ABJM duality.

\section{Supersymmetric D4-brane action on fuzzy $S^2$}\label{susyD4}

We will now build upon the results obtained in \cite{Nastase:2009ny} for the bosonic part of the action for fluctuations around the irreducible vacuum of the mass-deformed ABJM theory of \cite{Gomis:2008vc}. This was given by
\begin{multline}\label{initaction}
S^B=\int d^3x d^2\sigma \;\sqrt{ \hat h} 
 \Big[-\left(\frac{k}{2\pi}\right)^2\frac{ 1}{8 f^2} F_{\mu\nu}
 F^{\mu\nu} -\frac{N^2 \mu^2}{2}F_{ab}F^{ab} -\frac{N^2}{4}\partial
_\mu A^a \partial^\mu A_a  +Nf \d_\mu A_a\d^a A_\mu\\
-f^2 \d^a A_\mu \d_a A^\mu  + N^2 \mu^2 F_{ab}\hat \omega^{ab}\Phi  - \frac{N^2}{4} \partial_\mu  
\Phi \partial^\mu \Phi- N^2 \mu^2 \d_a \Phi \d^a \Phi- N^2 \mu^2 \Phi^2  \\
-4N^2 \mu^2 \d^a q_{\dot \a }^\dagger \d_a q^{\dot \a } - N^2  \d^\mu q_{\dot \a }^\dagger \d_\mu q^{\dot \a }\Big]\;.
\end{multline}
We extend the analysis to include fermions and find the full supersymmetric action, with and without the twisting of certain fields due to the partly compactified nature of the higher-dimensional worldvolume.  We also comment on the similarities and differences between the fuzzy sphere vacua of massive ABJM and the fuzzy funnel solutions of pure ABJM.

\subsection{Expectations from supersymmetry}

Although in the following we will focus our attention on the fuzzy sphere solution of the mass-deformed ABJM model, we will also comment on the fuzzy funnel solution of pure ABJM. In order to see what the expected result should be, we will first analyse the supersymmetry of the solutions.  We will use results already derived for the case of the BLG model and its massive deformation \cite{Bagger:2007vi,Gomis:2008cv,Hosomichi:2008jb},\footnote{In the context of the BLG theory, the M2$\perp$M5 intersection was investigated in \cite{Bagger:2007vi,Krishnan:2008zm}.} which suffice for our purposes. The BLG $\cA_4$-theory corresponds to an $\SU(2)\times \SU(2)$ ABJM model, which shares many qualitative features with the $\U(N)\times \U(N)$ constructions.

The ${\cal N}=8$ (\ie 16 supercharges in 3d) linearly realised supersymmetries of the massive deformation of BLG are given by \cite{Gomis:2008cv}
\bea
\delta_l X^I &=& i\bar\epsilon \Gamma^I\Psi\cr
\delta_l \Psi&=&D_\mu X^I \Gamma^\mu \Gamma^I \epsilon -\frac{1}{6}[X^I,X^J,X^k]\Gamma^{IJK}\epsilon-\mu
\Gamma_{3456}\Gamma^I X^I\epsilon \cr
\delta_l {{A_\mu}^b}_a&=&i\bar \epsilon \Gamma_I X^I_c \Psi_d{f^{cdb}}_a\;.
\eea
Here $I=1,...,8$ and the 3d Majorana spinor $\epsilon$ satisfies $\Gamma_{012}\epsilon=\epsilon$. These 
transformation rules are explicitly $\SO(8)$ invariant, as needed for ${\cal N}=8$ supersymmetry in 3d. However, 
the massive deformation, and in particular its vacuum solution, explicitly breaks the $\SO(8)$ symmetry of the 
action down to 
an $\SO(4)\times \SO(4)$ R-symmetry. Splitting $I$ into $4+4$ as $(A,A')$, the vacua of the mass-deformed theory 
are
\be
[X^A,X^B,X^C]=-\mu \epsilon^{ABCD}X^D;\;\;\; X^{A'}=0;\;\;\; \Psi=A_\mu=0
\ee
plus  the ones with $A$ and $A'$ indices interchanged. It is easy to check that on this solution
$\delta_l\Psi=\delta_l X^I=\delta_l {{A_\mu}^b}_a=0$, so all the 16 supercharges are preserved.

On the other hand, at $\mu=0$ (pure ABJM), the BPS fuzzy funnel solution\;, 
\be
\d_s X^A=\epsilon^{ABCD} [X^A,X^B,X^C]\;,\qquad X^{A'}=0\;,\qquad \Psi=A_\mu=0\;,
\ee
can be easily checked to imply $\delta_l\Psi=\delta_l X^I=\delta_l {{A_\mu}^b}_a=0$ only if
\be
\Gamma_s\epsilon=\Gamma_{3456}\epsilon\;,
\ee
\ie only $\frac{1}{2}$ of the linearly realised supersymmetries, namely 8 supercharges, are preserved.

A similar behaviour is also observed in the ABJM case, with all of 12 supercharges (${\cal N}=6$) surviving for the fuzzy sphere vacuum solution of massive ABJM, but only 6 supercharges (${\cal N}=3$) for the fuzzy funnel solution of pure ABJM. The correct effective action for small fluctuations around these classical solutions is expected to preserve the same number of supersymmetries.

Note that in \cite{Nastase:2009ny} it was shown that the (unrescaled) large-$N$ bosonic action is the same for both the fuzzy sphere and the fuzzy funnel solutions, so this difference in the number of preserved supersymmetries presents a  puzzle. We will return to this issue at the end of this section, where we will see that the fermionic part as well as the rescaling of the action imply the need for extra degrees of freedom to be taken into account in the fuzzy funnel case.

\subsection{Maximally supersymmetric D4 action}

The definition of the bosonic D4-brane fields, coming from the quadratic fluctuation action around the fuzzy sphere background of massive ABJM, was obtained straightforwardly in \cite{Nastase:2009ny} with one notable exception: The scalar fields $q^{\dot\a}$, which were overall transverse to both the worldvolume and the emergent $S^2$, were bifundamental, thus at finite $N$ they had to be expanded in terms of the bifundamental objects $\tilde{G}^\a Y_{lm}(J_i)$. However, since in the classical limit the $\tilde{G}^\a$ become Killing spinors on the sphere, this dependence alone suggests a spinor structure for said scalars.

Note that the appearance of the Killing spinor on $S^2$ as the classical limit of some fuzzy object is a feature that has not been previously considered in the literature. In the conventional construction of fermions on the fuzzy sphere, one obtains them as scalar functions expanded in terms of the usual spherical harmonics $Y_{lm}(J_i)$. The fermionic structure of the field is then obtained by diagonalising the kinetic operator in the classical limit.

For $q^{\dot\a}=Q_\a^{\dot\a}\tilde{G}^\a$ however, the action in \cite{Nastase:2009ny} corresponded to a usual scalar despite the presence of the fuzzy Killing spinor,  \eg one had that  the kinetic term on the sphere was given in terms of the familiar form 
\be\label{fullquadratic}
 \sim \mu^2N \int d^3 x d^2\sigma \sqrt{\hat h}\; \hat h^{ab}\;\d_a q^\dagger_{\dot \a} \d_b q^{\dot \a}\;,
\ee
even though $q^{\dot\a}$ `contains' the  Killing spinor $\tilde{G}^\a$ dependence. This is as long as one keeps in mind the following definition for the action of the  derivative on the Killing spinors  
\be
\d_a (\tilde{G}^\a)=\frac{1}{-2i}\hat h_{ab}K_i^b {(\ts_i)^\a}_\b \tilde{G}^\b\;.\label{derkill}
\ee
These two different possibilities in expressing the transverse scalar degrees of freedom are related to an implicit twisting, since $\tilde{G}^\a$ can be reinterpreted as part of the spherical harmonics. We will deal with this in detail in the next subsection, when we will twist the field $q^{\dot\a}$ into the spinor $Q_\a^{\dot\a}$ by `extracting' the $\tilde{G}^\a$. In this subsection we will instead turn to the fermions.

\subsubsection{Parallel fermions}

To complete the fluctuation action of \cite{Nastase:2009ny}, we begin with the fermionic part of the mass-deformed ABJM action 
\begin{multline}
S^f_{\mathrm{ABJM}}=\int d^3 x\; \Tr\Big[ - \psi^{I\dagger}\gamma^\mu D_\mu \psi_I  - i \mu \psi^{\dagger \alpha}\psi_\alpha + \frac{2\pi i}{k} \Big(\epsilon^{IJKL}\psi_IC^\dagger_J\psi_KC^\dagger _L-\epsilon_{IJKL}\psi^{\dagger I}C^J
\psi^{\dagger K}C^L\\
+C^\dagger_IC^I\psi^{J\dagger}\psi_J-\psi^{\dagger J}C^IC^\dagger_I
\psi_J-2 C^\dagger_IC^J\psi^{\dagger I}\psi_J+2\psi^{\dagger                                                                                     J}C^IC^\dagger_J\psi_I\Big)\Big]
\label{abjmfermions}\;,
\end{multline}
which differs from the undeformed ABJM fermionic action only in the presence of the mass term. Here $\psi_I$ are general (Dirac) spinors of $\SO(2,1)$, with 16 real components
and 8 on-shell degrees of freedom.
One can then split the fermions into two types, in a similar fashion to what we did for the scalars: the `parallel' to the $S^2$, which we denote as $\psi_\a$ and `transverse', which we will call $\chi_{\dot{\a}}$.  

For the parallel fermions $\psi_\a$, the terms with $\epsilon_{IJKL}$ in (\ref{abjmfermions}) do not contribute in the fuzzy sphere background.  The terms on the last line of (\ref{abjmfermions}) give
\be
\left(\frac{2\pi f^2}{k}\right)2i \Tr\left[ \frac{1}{2}(\bar{J}-J)\psi^{\dagger\a}\psi_\a +(\psi^{\dagger\b}{J^\a}_\b 
-{\bar{J}_\b \,}^\a \psi^{\dagger\b})\psi_\a\right]\;.
\ee

The fermions are also bifundamentals like the scalars $q^\a$, thus when expanding them in spherical harmonics we must
also consider a $\tilde{G}^\a$ multiplying the $Y_{lm}(J_i)$, as argued in \cite{Nastase:2009ny}. This leads to the natural decomposition
\bea
\psi_\a&=&  \tilde{G}^\a \psi+\tilde{G}^\b {\tilde{U}_\b\,}^\a\nonumber\\
&=& \tilde{\psi}\tilde{G}^\a+{U_\a}^\b \tilde{G}^\b\;,
\eea
where $\psi_\a$, ${\tilde{U}_\b\,}^\a$ admit an expansion in terms of $Y_{lm}(\bar J_i)$, while $\tilde{\psi}_\a, {U_\a}^\b$ in $Y_{lm}(J_i)$.  We also define raising and lowering of the indices on $\tilde{G}^\a$ by
\be
\tilde{G}_\a=\epsilon_{\a\b}\tilde{G}^\b;\;\;\;
\tilde{G}^{\dagger \a}=\epsilon^{\a\b}\tilde{G}^\dagger_\b\;.
\ee
Here
\be\label{raising}
\epsilon^{\alpha\beta}= \twobytwo{0}{1}{-1}{0}\;,
\ee
with $\epsilon^{\alpha\beta}\epsilon_{\beta\gamma} = \delta^\alpha_\gamma$, \ie as matrices $\epsilon = i \tilde \sigma_2$ and $\epsilon^{-1} = -i \tilde \sigma_2 $. Note also that $ \tilde{G}^\dagger_\alpha \tilde{G}^\alpha= -\tilde{G}^{\dagger\alpha}\tilde{G}_\alpha $.

Since
\bea
(\psi^{\dagger\b}{J^\a}_\b -{\bar{J}_\b \,}^\a \psi^{\dagger\b})\psi_\a&=&
\frac{J-\bar J}{2}\psi^{\dagger\a}\psi_\a+\frac{1}{2}{(\ts_i)^\a}_\b [\psi^{\dagger\b}J_i-\bar J_i\psi^{\dagger\b}]\psi_\a,
\eea
the kinetic term for $\psi_\a$ is ($2\pi f^2=\mu k$)
\be
i\mu \int \Tr {(\ts_i)^\a}_\b [\psi^{\dagger\b}J_i-\bar J_i\psi^{\dagger\b}]\psi_\a\;.
\ee

In \cite{Nastase:2009ny} it was shown that derivations on the sphere for fields with a  $Y_{lm}(J_i)\tilde{G}^\a$  dependence were obtained by considering
\be
q^\dagger_{\dot{\beta}}J_i-\bar J_i q^\dagger_{\dot{\b}}\rightarrow
+2i K_i^a \d_a q^\dagger_{\dot{\b}}+q^\dagger_{\dot\b}x_i\;.
\ee
Similarly, we now obtain 
\be
[\psi^{\dagger\b}J_i-\bar J_i\psi^{\dagger\b}]\rightarrow
+2iK_i^a\d_a \psi^{\dagger \b}+\psi^{\dagger\b}x_i\;.
\ee

The spin-covariant derivative on the sphere is given by \cite{Andrews:2006aw}
\be
\nabla_a = \d_a  + \frac{i}{4} \omega_a^{ij} \sigma_{ij}\;,
\ee
with $\omega^{ij}_a$ the spin connection on $S^2$, with only non-zero component $\omega_\phi^{12} = -\omega_\phi^{21}=-\cos{\theta}$,
and $\sigma_{ij}$ the spin-$\frac{1}{2}$ generators of SO(2). In terms of components
\be\label{components}
\nabla_1\equiv \nabla_\theta=\d_\theta\;,\qquad
\nabla_2\equiv \nabla_\phi=\d_\phi -\frac{i}{2}\Gamma_3\cos\theta
\ee
and one can write the above as 
\be
\nabla_a = \d_a + S^{-1}\d_a S +\frac{i}{2}\Gamma_a \label{diracs}\;.
\ee
Hence, the Dirac operator on the sphere is given by 
\be
\tilde{D}_2\equiv e^a_m\Gamma^m\nabla_a
=\sigma_1\Big( \d_\theta+\frac{\cot\theta}{2}\Big)+\sigma_2\frac{1}{\sin\theta}\d_\phi
=\Gamma^a(\d_a +S^{-1}\d_a S)+i\;,
\ee
where $S$ is same unitary rotation matrix previously  defined in (\ref{Smatrix}), which also appears when translating quantities on $S^2$ between Cartesian and spherical coordinates. We have collected definitions and various identities involving the matrices $S$ in Appendix\;\ref{AppA}.

Using (\ref{gamma3}) and (\ref{diracs}), we get the following kinetic term
for  $\psi_\a$,  coming from the $CC\psi\psi$ interaction term 
\be\label{28}
-2\mu N  \int d^3 x d^2 \sigma  \sqrt{\hat h}\; \Big[(\psi S)_\a{[\Gamma^a  \nabla_a -i P_+]^\a}_\b
(S^{-1}\psi^{\dagger})^{ \b})\Big]\;,
\ee
where again the projector $P_\pm = \frac{1}{2}(1\pm\Gamma_3)$.

To this, we must add the 3d kinetic term for these parallel fermions plus the mass term coming from the deformation
\be
-\int \Tr[\psi^{\dagger\a}\gamma^\mu \pd_\mu \psi_\a +i\mu \psi^{\dagger\a}\psi_\a] 
\rightarrow-N\int d^3 x d^2 \sigma\sqrt{\hat h}\;[\psi^{\dagger\a}\gamma^\mu \pd_\mu \psi_\a +i\mu \psi^{\dagger\a}\psi_\a]\;,
\ee
where in the above the covariant derivative drops out because the $\psi A\psi$ interaction terms are cubic in the fluctuating fields.

We total action for the parallel fermions $\psi_\a$ is then
\be
N\int d^3 x d^2 \sigma\sqrt{\hat h}\;
(\psi S)_\a{[-\gamma^\mu \d_\mu \one+2\mu(-\Gamma^a\nabla_a+\frac{i}{2} \Gamma^3)]^\a}_\b
(S^{-1}\psi^{\dagger})^\b\;.
 \label{paral}
\ee

This is almost the kinetic term of a 5d fermion on an $S^2$ of radius $\frac{1}{2\mu}$. Indeed, there is a unique split of the $\Gamma$-matrices in 5d into 3d+2d, namely $\hat \Gamma^\mu=\gamma^\mu \otimes \Gamma^3$, $\hat \Gamma^a=1\otimes \Gamma^a$, so the 5d Dirac operator must be
\be\label{dirac5d}
D_5=\hat \Gamma^\mu \d_\mu +\hat \Gamma^a 2\mu \nabla_a=
\gamma^\mu {(\Gamma_3)^\a}_\b\d_\mu+2\mu {(\Gamma^a)^\a}_\b \nabla_a\;.
\ee
Note that what is  missing is a Weyl condition, \ie if we had Weyl spinors, with
$(1-\Gamma_3)\psi=0$, or $\frac{1+\Gamma_3}{2}\psi=\psi$,
we would get the above result in terms of the 5d Dirac operator in (\ref{paral}).

The D4-brane action that we want to finally obtain, should sport a Majorana spinor in 5d. However this decomposes into a Majorana spinor in 3d, times a Weyl spinor or a Majorana spinor in 2 Euclidean dimensions.  It should also come with an index for the 4 dimensional real representation of the D4-brane R-symmetry group $\SO(5)_R$. Since the $\psi_\a$ correspond to half the number of the total D4-brane fermions (the others being related to $\chi_{\dot \alpha}$), one still needs an extra index $i=1,2$ on the 5d fermion, or equivalently to have a Dirac spinor in 5d instead of Majorana.

From the point of view of the lower dimensional theory we started with a general (complex Dirac) spinor in 3d. It is then clear that to obtain a complex Dirac spinor in 5d from the fuzzy sphere we must have a Weyl spinor on the 2-sphere.  Thus the subtlety is that, by interpreting the index $\a=1,2$ on $\psi_\a$ as an index on the fuzzy 2-sphere, we must reorganise it as a 2d-Weyl spinor index, \ie we must impose a Weyl condition.  Thus the need for the Weyl condition appears when comparing degrees of freedom at finite $N$ and on the classical 2-sphere, and is related to the presence of the strange object $\tilde{G}^\a$ in the decomposition of the fields. Indeed, we saw that $\tilde{G}^\a$ corresponds to the Weyl-projected Killing spinor $P_+\eta^I$, where $P_+$ acts either on the $\SO(2)_L$ index or on the $\SO(3)_{global}$ index $I$.  We will see in the next section that if we take out the $\tilde{G}^\a$, we obtain the Weyl projection automatically, without the need to impose it by hand. Also, when twisting $\psi$ by removing a $\tilde{G}^\a$ in the next subsection, this kind of subtlety will disappear.

In conclusion, we obtain a 5d spinor $\psi$ that is 2d-Weyl, with mass $\mu$.

\subsubsection{Transverse fermions}\label{transferm}

We now move to the transverse fermions $\chi_{\dot{\a}}$. From the $\epsilon^{IJKL}$
term in (\ref{abjmfermions}) one has
\be\label{first}
i\mu\; \Tr\left[ \chi^{\dagger\dot{\a}}\chi_{\dot{\a}}
- \epsilon^{\alpha\gamma}\epsilon^{\dot\beta\dot{\delta}}\tilde{G}^\dagger_\a 
\chi_{\dot{\b}}\tilde{G}^\dagger_\g\chi_{\dot{\delta}}
+\epsilon_{\alpha\gamma}\epsilon_{\dot{\beta}\dot{\delta}}\tilde{G}^\a\chi^{\dagger\dot{\b}}\tilde{G}^\g \chi^{\dagger\dot{\delta}}\right]\;.
\ee
As for the parallel fermions, $\chi_{\dot\a}$ are bifundamentals so we must extract a $\tilde G^\b$ matrix before decomposing in terms of the  fuzzy spherical harmonics $Y_{lm}(J_i)$
\bea
\chi_{\dot\a}&=&\chi_{\dot\a \b}\tilde{G}^\b\cr
\chi^{\dagger\dot \alpha}&=&\tilde{G}^\dagger_\b\chi^{\dot\a \b}\;.
\eea
The dotted indices are raised and lowered in the same way as the undotted indices. Note that the above implies the modified Majorana spinor condition 
\be
(\chi_{\dot\a \a})^\dagger=\chi^{\dot\a \a}\label{realit}\;.
\ee
This is needed, since the fields $\chi_{\dot\a}$ were general (complex Dirac) spinors in 3d, but by 
extracting $\tilde{G}^\a$, the $\chi_{\dot\a \a}$ need to obey a reality condition. 

After some algebra one obtains for the two nontrivial terms in (\ref{first})
\be
-\frac{\mu i}{4}(2i)\epsilon_{jik}
\Tr\left[{(\ts_k)^\a}_\b J_j \chi^{\dot{\delta}\beta} J_i\chi_{\dot{\delta}\alpha}\right]\;.
\ee
The expression inside the bracket gives in the classical limit
\be
\epsilon_{jik}J_j \chi^{\dot{\delta}\beta} J_i=-\epsilon_{jik}J_j[J_i, \chi^{\dot{\delta}\beta}]
+2iJ_k\chi^{\dot{\delta}\beta}\;\;\;\;
\rightarrow \;\;\;\; [-N\epsilon_{jik}x_j(-2i)K_i^a\d_a+2Ni x_k]\chi^{\dot{\delta}\b}\;,
\ee
which through use of the identity
\be
\epsilon_{ijk}x_iK_j^a=\hat \omega^{ad}\hat h_{dc}K_k^c
\ee
 gives
\be
\epsilon_{jik}J_j \chi^{\dot{\delta}\beta} J_i \;\;\;\; \rightarrow \;\;\;\; 2iN[\hat \omega^{ad}\hat h_{dc}K_k^c\d_a+x_k]\chi^{\dot{\delta}\beta}\;.
\ee
Using (\ref{diracs}), the identities  (\ref{SKx}), as well as the relations
\bea
&&\hat\omega^{ad}\Gamma_d\nabla_a=-i\Gamma_3\Gamma^a\nabla_a\cr
&&\hat\omega^{ad}\Gamma_d\Gamma_a = -2i\Gamma_3\;,
\eea
which can be checked by explicit evaluation, one eventually arrives at the following result 
\be
\rightarrow
 -N^2\int d^3x d^2 \sigma \sqrt{\hat h}\; \Big[\mu(\chi_{\dot{\delta}}S)_\alpha \Big( P_+  {\Gamma^a \nabla_a \Big)^\alpha}_\beta {(S^{-1}\chi^{\dot{\delta}}})^\beta +h.c.\Big]\;.
\ee

Of course, one  also needs to add the usual kinetic and mass terms in 3d for $\chi_{\dot\a}$ (see (\ref{abjmfermions})), namely
\be
\int \Tr[-\chi^{\dagger\dot\a}\gamma^\mu D_\mu \chi_{\dot\a} +i\mu \chi^{\dagger\dot\a}\chi_{\dot\a}] 
\rightarrow N\int d^3x d^2 \sigma \sqrt{\hat h}\;[- \chi^{\dagger\dot\a}\gamma^\mu \pd_\mu \chi_{\dot\a} +i \mu \chi^{\dagger\dot\a}\chi_{\dot\a}]\;.
\ee
Combining that with the mass-term that has been left over from (\ref{first}) one gets
\be
\rightarrow N\int d^3x d^2 \sigma \sqrt{\hat h}\;[- \chi^{\dagger\dot\a}\gamma^\mu \pd_\mu \chi_{\dot\a} +2i \mu \chi^{\dagger\dot\a}\chi_{\dot\a}]\;.
\ee
By expressing the above in terms of $\chi_{\dot\a \a}$, using  ${J^\a}_\b\rightarrow \frac{N}{2}(x_k{(\ts_k)^\a}_\b+\delta^\a_\b)$, this is 
\be
N^2\int d^3x d^2 \sigma \sqrt{\hat h}\;\Big[-(\chi^{\dot{\a}}S)_\b{(P_-)^\beta}_\alpha \slashed{\d}(S^{-1}\chi_{\dot{\a}})^\alpha + 2i\mu (\chi^{\dot{\a}}S)_\b{(P_-)^\beta}_\alpha (S^{-1}\chi_{\dot{\a}})^\alpha\Big]\;,
\ee
where $\slashed{\d} = \gamma^\mu \d_\mu$ as usual, and the total action for the transverse fermions is 
\begin{multline}
N^2\int d^3x d^2 \sigma \sqrt{\hat h}\;\Big[\frac{1}{2}(\chi^{\dot{\a}}SP_-)_\b{(\Gamma^3)^\beta}_\alpha \slashed{\d}(P_-S^{-1}\chi_{\dot{\a}})^\alpha + i\mu (\chi^{\dot{\a}}SP_-)_\alpha (P_-S^{-1}\chi_{\dot{\a}})^\alpha\\+\mu (\chi^{\dot{\delta}}S  P_+)_\alpha \Big(  {\Gamma^a \nabla_a 
\Big)^\alpha}_\beta {(P_-S^{-1}\chi_{\dot{\delta}}})^\beta +h.c.\Big]\;.
\end{multline}
Here we introduced a $\Gamma^3$ in front of $\slashed{\d}$ in order to make explicit the correct decomposition of the 
5d Dirac spinor. 

Now we can define 
\be
\Upsilon_{\dot\a}^\a={(P_-S^{-1}\chi_{\dot{\a}}})^\a
\ee
and, as promised, the Weyl projection $P_-$ appears automatically, for the same reasons as mentioned for the parallel fermions: for the counting of degrees of freedom to work  one needs to construct either a single 5d Dirac fermion or two 5d Majorana fermions.  In this case, the appearance of the Hermitian conjugate means that one must ignore the (modified) Majorana reality condition. Alternatively, one could reorganise the spinors into (modified) Majorana spinors but without the Weyl condition, as the two results are equivalent. We will not do this here, although we will perform the equivalent procedure when twisting the transverse scalars $q^{\dot\a}$ shortly.

In terms of the $\Upsilon^\alpha_{\dot \alpha}$'s the action for the 
transverse fermions is 
\be\label{transversefer}
N^2\int d^3x d^2 \sigma \sqrt{\hat h}\;\Big[\frac{1}{2}\bar \Upsilon^{\dot \alpha} D_5  \Upsilon_{\dot\alpha} + i \mu \bar \Upsilon^{\dot\alpha} \Upsilon_{\dot{\a}} +h.c.\Big]\;.
\ee
Here we have also used the 5d Dirac operator (\ref{dirac5d}) that includes a sphere factor of radius $\frac{1}{2\mu}$. 

In conclusion, the spinor $\Upsilon_{\dot\a}^\a$ has a Weyl-projected sphere index $\a$, making it the expected D4-brane Dirac fermion.

\subsubsection{Final action and supersymmetry}

Collecting all contributions, the action will become just the usual D4-brane action for bosonic fields $\Phi, q^{\dot\a}, A_\mu, A_a$ and fermionic fields 
$\psi_\a, \Upsilon_{\dot\a}$, but with $q^{\dot\a}$ and $\psi_\a$ `containing' a fuzzy Killing spinor. We will see shortly that this can bee interpreted in terms of a twisting of these fields. The action is
\begin{multline}\label{intermedaction}
S=\int d^3x d^2\sigma \;\sqrt{ \hat h} 
 \Big[-\left(\frac{k}{2\pi}\right)^2\frac{ 1}{8 f^2} F_{\mu\nu}
 F^{\mu\nu} -\frac{N^2 \mu^2}{2}F_{ab}F^{ab} -\frac{N^2}{4}\partial
_\mu A^a \partial^\mu A_a  +Nf \d_\mu A_a\d^a A_\mu\\
-f^2 \d^a A_\mu \d_a A^\mu -4N^2 \mu^2 \d^a q_{\dot \a }^\dagger \d_a q^{\dot \a } - N^2  \d^\mu q_{\dot \a }^\dagger \d_\mu q^{\dot \a }    - \frac{N^2}{4} \partial_\mu  
\Phi \partial^\mu \Phi- N^2 \mu^2 \d_a \Phi \d^a \Phi- N^2 \mu^2 \Phi^2  \\
+ N^2 \mu^2 F_{ab}\hat\omega^{ab}\Phi +N^2\Big(\frac{1}{2}\bar \Upsilon^{\dot \alpha} D_5  \Upsilon_{\dot\alpha} + i \mu \bar \Upsilon^{\dot\alpha} \Upsilon_{\dot{\a}} +h.c.\Big)+N\Big((\psi S)_\a{[-D_5+i\mu \one]^\a}_\b (S^{-1}\psi^\dagger)^\b\Big)
\Big]\;.
\end{multline}
Note that we have already assumed that the $\psi$ fermions are Weyl-projected. 

As in \cite{Nastase:2009ny}, in order to bring the above to a form that can be compared to a conventional D-brane action, it is necessary to redefine the matter fields by $C^I\to X^I = (T_2^{-1/2} f G^\alpha,0) $ and hence $f\to T_2^{-1/2} f$, where $T_2 = [l_p^3 (2\pi)^2]^{-1}$ the membrane tension, and similarly for the fermions. This is so that the $X^I$'s can be thought of as spacetime coordinates with dimensions of length. We then perform some additional rescalings for the bosonic fields 
\be
 A_\mu \to
 A_\mu\, \frac{4 \pi l_s }{T^{-1/2}_{2}f}\;,\quad A_a \to A_a \,\frac{4 \pi l_s}{N} \;,\quad \Phi\to
\Phi \,\frac{4 \pi l_s}{N\mu} \;,\quad q^{\dot \alpha} \to q^{\dot \alpha} \,\frac{ 4 \pi l_s}{\sqrt{N}\mu}\;,
\label{rescabos}
\ee
for the sphere metric $ h_{ab} = \mu^{-2} \hat h_{ab} $ and the worldvolume coordinates $x^\mu\to \frac{1}{ 2} x^\mu$. These are finally supplemented  by the following rescalings of the fermions
\be
\Upsilon^{\dot\a} \to \Upsilon^{\dot\a} \frac{4\pi l_s}{N\mu}\;, \qquad
\psi_\a\to \psi_\a \frac{4\pi l_s}{\sqrt N\mu}\;.
\label{rescaferm}
\ee
After implementing the above, we arrive at 
\bea\label{rescaled}
S_{phys} &=& \frac{1}{g_{YM}^2}\int d^3 x d^2 \s \sqrt { h }\; 
 \Big[-\frac{1}{4}  F_{AB}  F^{AB}-\frac{1}{2} \partial_A 
\Phi \partial^A\Phi - \frac{\mu^2}{2} \Phi^2 -\d^M q_{\dot\a}^\dagger \d_M q^{\dot\a}
+ \frac{\mu}{2}\; \omega^{ab} F_{ab}\Phi\nn\\
&&\qquad\qquad +\Big(\frac{1}{2}\bar \Upsilon^{\dot \alpha} \tilde D_5  \Upsilon_{\dot\alpha} + 
\frac{i}{2} \mu \bar \Upsilon^{\dot\alpha} \Upsilon_{\dot{\a}}+h.c.\Big)
-(\psi S) \tilde D_5 (S^{-1}\psi^\dagger) +\frac{i}{2}\mu (\psi S) (S^{-1}\psi^\dagger) \Big]\;,\nn\\
\eea
where $A_M = \{A_\mu, A_a\}$, $\tilde D_5= \gamma^\mu {(\Gamma_3)^\a}_\b\d_\mu+\mu {(\Gamma_a)^\a}_\b\nabla^a$. This is just the action of a partly spherical D4-brane  with some extra mass terms and 12 supercharges on the worldvolume, or twice as much in the curved spacetime background! The mass terms break $\SO(4,1)$ Lorentz invariance, which is not that surprising as the  worldvolume itself already breaks it. They are also separately maximally supersymmetric from the point of view of 3d.\footnote{One can easily check that $\delta F_{ab}\propto   \mu\omega_{ab}\bar\epsilon\psi,\delta \phi\propto \bar\epsilon \psi, \delta \psi   \propto \mu\phi\epsilon$ with $\psi$ a generic fermion leave the mass terms invariant.} We will not attempt to make the full supersymmetry transformations explicit here, as they will be of a peculiar type, but will instead focus on their general characteristics. We will soon explain in more detail why we must obtain 12 supercharges, but the D-brane action in curved space must preserve $\frac{1}{2}$ of the supersymmetry of the background.

We now recall how we expanded the various ABJM fields in the classical limit of the sphere. For the adjoint gauge fields it was done in the usual manner in terms of scalar spherical harmonics, \ie $A_\mu^{(i)}=(A_\mu^{(i)})_{lm} Y_{lm}(x_i)$, while all other bifundamental fields were expanded in $Y_{lm}(x_i)\tilde{g}^\a$ as
\bea
r^\a&=&r\tilde{g}^\a+{s^\a}_\b \tilde{g}^\b=\Big[(r)_{lm}\delta^\a_\b+({s^\a}_\b)_{lm}\Big]Y_{lm}(x_i)\tilde{g}^\b\cr
q^{\dot\a}&=&Q^{\dot\a}_\a\tilde{g}^\a=(Q^{\dot\a}_\a)_{lm}Y_{lm}(x_i)\tilde{g}^\a\cr
\psi_\a&=&\tilde{\psi}\tilde{g}_\a+{U_\a}^\b \tilde{g}_\b=\Big[(\tilde{\psi})_{lm}\delta_\a^\b+({U_\a}^\b)_{lm}\Big]
Y_{lm}(x_i)\tilde{g}_\b\cr
\chi_{\dot\a}&=&\chi_{\dot\a \a}\tilde{g}^\a=(\chi_{\dot\a \a})_{lm}Y_{lm}(x_i)\tilde{g}^\a\;.\label{expansi}
\eea

Simply because of the form of the 2+1 dimensional part of the action, it is natural to expect that $r^\a$ must be  bosonic and $\chi_{\dot\a}$ must be fermionic. One can get from the initial mass-deformed ABJM action to the final result (\ref{rescaled}) through replacing $r_\a$ with the bosonic fields $A_a$ and $\Phi=2r+\phi$ (where ${s^\a}_\b {(\ts_i)^\b}_\a=K_i^aA_a+x_i\phi$) and $\chi_{\dot\a}$ with $\Upsilon_{\dot\a}^\a=(P_-S^{-1}\chi_{\dot\a})^\a$. 

In the same expression (\ref{rescaled}), $q^{\dot\a}$ and $\psi_\a$ were left as they were, since similarly the 2+1 dimensional part of the action implies that they are bosonic and fermionic fields respectively. Even though the form of the final expressions is extremely simple, this presents a kind of asymmetry in the way we have treated the fields, as the $q^{\dot \alpha}, \chi_{\dot \alpha}$ still contain a (fuzzy) Killing spinor in their expansion. The reader might be  wondering why we have also not naturally replaced $q^{\dot\a}$ with $Q^{\dot\a}_\a$ and $\psi_\a$ with $\tilde{\psi}, {U_\a}^\b$.  We will see in the next section that this will correspond to twisting the fields, which will turn $\tilde{\psi}, {U_\a}^\b$ into a combination of twisted-scalars and vectors, while $Q^{\dot\a}_\a$ into twisted-spinors.

An intriguing feature of this result is the following: while in the finite-$N$ construction $Q^{\dot\a}_\a$ and $\chi_{\dot\a \a}$ can only be expanded in $Y_{lm}(J_i)$, the fact that these fields are (twisted) spinorial means that, on the classical $S^2$, one should actually expand them in terms of  spinor spherical harmonics, \ie
\bea
&&Q^{\dot\a}_\a=\sum_{lm,\pm} (Q^{\dot\a})_{lm,\pm} \Xi^{\pm\a}_{lm}\cr
&&\chi_{\dot\a \a}=\sum_{lm, \pm} (\chi_{\dot\a})_{lm,\pm} \Xi^{\pm\a}_{lm}\;,
\eea
with $\Xi^{\pm\a}_{lm}$ as given in (\ref{killspsphhar}) and also containing the Killing spinor. Hence, the expansion (\ref{expansi}) must somehow rearrange itself at large-$N$. In other words, and according to the new construction presented in this paper, in the classical limit a spinor index can arise both from the $\tilde{G}^\a$ acting as a (fuzzy) spherical harmonic or (fuzzy) Killing spinor, and also from the coefficients of the expansion in fuzzy spherical harmonics, as it is usually done.

This unusual behaviour is related to the fact that in the classical limit, $\tilde{G}^\a$ matches against an object with 2 spinor indices, global and local, either one of which can be thought of as being removed by a Weyl projection as was discussed under Eq.\;(\ref{identindic}). By the finite-dimensional matrix rules, the bifundamental matrix $q^{\dot\a}$ can only be expanded in $ Y_{lm}(J_i)\times \tilde{G}^\a$ and we can think of the $\a$ index on $\tilde{G}^\a$ as a global symmetry index. However, at large $N$ it also can be reinterpreted as a local Lorentz (spinor) index. Since $\eta^\alpha$ is contracted with the coefficient $(Q^{\dot\a}_\a)_{lm}$, the latter also becomes a spinor.

We conclude this section with a few comments on the action of supersymmetry. The set of ${\cal N}=6$ supersymmetry transformations in 3d and at finite $N$ include \cite{Gomis:2008vc}
\be
\delta (C^I)^{ij}=\bar \epsilon^{IJ} (\psi_J)^{ij}\;,\label{finiten}
\ee
where we have explicitly written the $\U(N)\times \U(\bar N)$ matrix $(ij)$ indices. This could be decomposed into
\be
\delta_1 C^{\dot\a}=\bar \epsilon_1^{\dot\a \a}\psi_\a\qquad \text{and}\qquad \delta_2 C^{\dot\a}=\bar \epsilon_2 ^{\dot\a\dot\b}
\chi_{\dot\b}\;,
\ee
 where $\epsilon^{IJ}$ is in the 6-dimensional, antisymmetric representation of $\SU(4)$.

At $N\rightarrow\infty$ one still has ${\cal N}=6$ supersymmetry.
In the classical supersymmetric D4-brane action (\ref{rescaled}), supersymmetry similarly relates
$q^{\dot\a}$ with $\psi_\a$ and  $\Upsilon^{\dot\a}_\a$. The first half of the (global) supersymmetry transformations
\be
\delta_1q^{\dot\a}=\bar{\epsilon}_1^{\dot\a \a}\psi_\a
\ee
is of the usual kind, since both $q^{\dot\a}$ and $\psi_\a$ are bifundamental matrices at finite $N$, and $\epsilon_1$
does not act on the matrix structure, as in the classical limit the Lorentz spinor index on $\epsilon_1$ naturally appears from (\ref{finiten}). On the other hand, the other half,
\be
\delta_2q^{\dot\a}=\bar \epsilon_2^\a\Upsilon^{\dot\a}_\a=\bar\epsilon_2\tilde{g}^\a
\Upsilon^{\dot\a}_\a\;,
\ee
has a more unusual supersymmetry parameter, since $q^{\dot\a}$ is bifundamental while $\Upsilon^{\dot\a}_\a$ is adjoint, so for this transformation to make sense away from infinite $N$, one must decompose $\epsilon_2^\a$ as above. That, however, would mean that supersymmetry would act on the gauge group and hence cannot originate from (\ref{finiten})!  In the classical limit,  $\epsilon_2^\a$ should of course be the same kind of object as $\epsilon_1$, a spinor on the sphere, but the consistency of the $N\to\infty$ limit must be subtle in order to obtain the correct supersymmetry from the finite-$N$ one.

It is apparent that if one replaced instead $q^{\dot\a}$ by $Q^{\dot\a}_\a$ and $\psi_\a$ by $\tilde{\psi}$, ${U_\a}^\b$ such a problem would be avoided and the classical limit would be better defined, since all the fields at finite $N$ are then in the adjoint of $\U(N)$, and can be treated on the same footing.

\subsection{Twisting the D4 action on the fuzzy $S^2$}

Following the above discussion, the alternative way of expressing the action for fluctuations around the irreducible vacuum is such that all the classical fields on the sphere admit an expansion in the scalar fuzzy spherical harmonics $Y_{lm}(J_i)$, with the spinorial structure of some fields appearing solely from the coefficients of that expansion. This is the natural construction for the fields on the fuzzy sphere but in this picture we will end up with a set of `twisted' fields, in a sense that we will shortly explain. This affects the transverse fermions $\chi_{\dot\a}$, as well as the expression for the transverse scalars $q^{\dot   \alpha}$ found in \cite{Nastase:2009ny}.

\subsubsection{Twisted Compactification vs. `Deconstruction' in the Maldacena-N\'u\~nez model}

We initially review the similar case of \cite{Andrews:2006aw}, in the context of the Maldacena-N\'u\~nez (MN) model with IIB 5-branes compactified on $S^2$, preserving ${\cal N}=1$ supersymmetry in 4 dimensions (the mass-deformed ${\cal N}=1^*$ theory of \cite{Polchinski:2000uf}).  As is known from \cite{Bershadsky:1995qy}, in order to preserve supersymmetry on D-branes with curved worldvolumes, one needs to twist the various D-brane fields. Specifically, that means embedding the $S^2$ spin connection, taking values in $\SO(2)\simeq\U(1)$, into the R-symmetry. As a result, the maximal supersymmetry one can obtain after compactification to 4 dimensions is ${\cal N}=1$ (corresponding to $\U(1)_R$), which the MN twisting indeed does result to. The authors of \cite{Andrews:2006aw} then compare the MN twisted compactification with a `deconstruction' of an ${\cal N}=1^*$, $\SU(N)$ gauge theory at large-$N$ and around a fuzzy $S^2$ background, obtaining agreement in the spectrum and action for fluctuations. We now turn to understanding this twisting procedure, in order to apply the lessons learnt to the case of the ABJM theory.

First note that there are two ways to understand the twisting: from the point of view of the twisted compactification on the sphere, as well as from the point of view of the `deconstruction' picture, \ie by constructing the fuzzy sphere from matrices in the lower dimensional theory, as we have performed so far.

On a 2-sphere, scalar fields are decomposed in the usual spherical harmonics $Y_{lm}(x_i)=Y_{lm}(\theta,\phi)$ and can thus give massless fields after compactification (specifically, the $l=0$ modes). However, that is not true any more for spinors and gauge fields.  Spinors on the sphere are eigenvectors of the total angular momentum on the sphere $J_i^2$. These are of two types: eigenvectors $\Omega$ of the orbital angular momentum $L_i^2$ (Cartesian spherical spinors) and eigenvectors $\Upsilon$ of the Dirac operator on the sphere $-i\hat{\nabla}_{S^2}=-i\hat h^{ab} e^{m}_a\sigma_m\nabla_b$  (spherical basis spinors), whose square is $R^2(-i\hat{\nabla}_{S^2})^2=J_i^2+\frac{1}{4}$. The two are related by a transformation with a sphere-dependent matrix $S$. The former are decomposed in the spinorial spherical harmonics
\be
\Omega^{\hat{\a}}_{jlm}=\sum_{\mu=\pm \sfrac{1}{2}}C(l,\sfrac{1}{2},j;m-\mu, \mu,m)Y_{l,m-\mu}(\theta,\phi)\chi_\mu^{\hat{\a}}\;,
\ee
where $j=q_{\pm}=l\pm \frac{1}{2}$ and $\hat{\a}=1,2$, as 
\be
\psi^{\hat\a}=\sum_{lm}\psi_{lm}^{(+)}\Omega_{l+\frac{1}{2},lm}^{\hat\a}+\psi_{lm}^{(-)} \Omega_{l-\frac{1}{2},lm}^{\hat\a}\;.
\ee
Both have a minimum mass of $\frac{1}{2R}$, since the Dirac operator squares to $J_i^2+\frac{1}{4}=j(j+1)+\frac{1}{4}$. Similarly, 
the vector fields do not simply decompose in $Y_{lm}$'s, but rather in the vector spherical harmonics
\bea
&& \frac{1}{R}{\bf T}_{jm}=\frac{1}{\sqrt{j(j+1)}}\Big[\sin\theta \d_\theta Y_{jm} {\bf \hat\phi}
-\csc\theta \d_\phi Y_{jm}{\bf\hat\theta}\Big]\cr
&& \frac{1}{R}{\bf S}_{jm}=\frac{1}{\sqrt{j(j+1)}}\Big[\d_\theta Y_{jm} {\bf \hat\theta}
+\d_\phi Y_{jm}{\bf\hat\phi}\Big]\;,
\eea
with $j\geq 1$. It is more enlightening to show the decomposition of the field strength on the 2-sphere,
\be
\frac{1}{R}\csc\theta F_{\theta\phi}=R^2\sum_{lm}F_{lm}\frac{1}{\sqrt{l(l+1)}}\Delta_{S^2}Y_{lm}\;,
\ee
with $l=1,2,...$, thus again only massive modes are obtained after dimensional reduction \cite{Andrews:2006aw}. 

Therefore, in the absence of twisting, supersymmetry will be lost after dimensional reduction, since all $S^2$-fermions will be massive but some massless $S^2$-scalars will still remain. Twisting, however, allows for the presence of fermionic twisted-scalars (T-scalars), \ie fermions that are scalars of the twisted $\SO(2)_T$ Lorentz invariance group (with charge $T$), which will stay massless, and the number of supersymmetries in the dimensionally reduced theory equals the number of fermionic T-scalars.

In compactifying a 5-brane on $S^2$, one has a $\SO(4)_R\simeq\SU(2)_A\times \SU(2)_B$ R-symmetry and a $\SO(3,1)\times \SO(2)_{45}$ (local) Lorentz invariance.  One chooses the twisted Lorentz invariance of the sphere as $Q_T=Q_{45}+Q_A$, where $Q_{45}$ is the charge under the original Lorentz invariance $\SO(2)_{45}$, and $Q_A$ is the charge under the $\U(1)$ subgroup of $\SU(2)_A$. The reason this is necessary is because one needs to identify the $\U(1)$ spin connection (`gauge field of Lorentz invariance') with a corresponding connection in an R-symmetry subgroup, \ie a gauge field from the transverse manifold. Note that $\SU(2)_B$ is inert (\ie unaffected by the sphere) and is thus a truly transverse group that can be called $\SU(2)_\perp$. The true symmetries of the twisted compactification are then $\SU(2)_{\perp}\times \U(1)_T$ and the usual Lorentz invariance $\SO(3,1)$.

The 5-brane bosonic fields are gauge fields $A_M$, 4 scalars $\phi^m$ charged under $\SO(4)_R\simeq\SU(2)_A\times \SU(2)_B$, with respective indices $\underline \alpha$ and $\underline{\dot \alpha}$. There are also two spinors, one charged under $\SU(2)_A$, $\lambda_l$, and one charged under $\SU(2)_B$, $\tilde{\lambda}_l $.  The twisted fields are the ones charged under $\SU(2)_A$, \ie $\phi^m$ and $\lambda_l$.  One writes $\phi^m=-\frac{i}{2}(\tau^m)^{\underline{\a}\underline{\dot     \a}}v_{\underline{\a}\underline{\dot\a}}$ showing explicitly the $\underline{\a}$ index of $\SU(2)_A$, and this field has twisted spin $Q_T=0+\frac{1}{2}=\frac{1}{2}$.\footnote{Note that we divide the charge used in \cite{Andrews:2006aw} by 2, preferring to keep the usual notation for spin over the   usual notation for U(1) charge.} For $\lambda_l$ one writes explicitly the Lorentz $\SO(3,1)\simeq\SU(2)_L\times \SU(2)_R$ and $\SU(2)_A$ indices, $\lambda ^\a_{\underline{\a}}$ and $\bar\lambda^{\dot\a}_{\underline{\a}}$, and decomposes $\frac{1}{2}\otimes \frac{1}{2}=0\oplus 1$ into a vector and a scalar, thus building the T-scalar ($Q_T=\frac{1}{2}-\frac{1}{2}=0$) $\Lambda$ from $\lambda^\a _{\underline{\a}=2}$, $\bar{\lambda}^{\dot{\a}}_{\underline{\a}=1}$ and the T-vector ($Q_T=\frac{1}{2}+\frac{1}{2}=1$) $g^a$ from $\lambda^\a_{\underline{\a}=1}$, $\bar{\lambda}^{\dot{\a}}_{\underline{\a}=2}$.  The untwisted fields comprise of $A_M$ splitting into bosonic T-scalars $A_\mu$ and bosonic T-vectors $A_a$, and the fermionic T-spinors.

The explicit form for the twisted fields (a bosonic T-spinor, fermionic T-scalars and T-vectors), is summarised in the action 
\be
\int \Big[ -\frac{i}{2}\mu \bar \Lambda \gamma^\mu \d_\mu \Lambda-\frac{i}{2}\mu
\bar g_a \gamma^\mu \d_\mu g^a+\mu \omega^{ab} \bar G_{ab}\Lambda-2\d_\mu \Xi^\dagger \d^\mu \Xi
-8\Xi^\dagger(-i\hat \nabla_{S^2})^2\Xi\Big]\;,\label{twistDorey}
\ee
where $\mu$ is the mass deformation parameter, $G_{ab}= \d_{a} g_{b}- \d_{b} g_{a}$ and as usual $\omega^{ab}=\frac{1}{\sqrt{g}}\epsilon^{ab}$ is the symplectic form on the sphere.  

We next try to understand why one needs to twist from the point of view of deconstruction, and why this leads to reproducing the same answer. In `deconstructing' the above action from 4d matrices (D3-branes), one has $\SO(3,1)$ Lorentz invariance and $\SO(6)_R\simeq \SU(4)_R$ R-symmetry, which is broken by the choice of fuzzy sphere solution to $ \SU(2)_1\times \SU(2)_2\times \U(1) \simeq \SO(3)\times \SO(3)\times \U(1)$, where the $\U(1)$ is a charge that rotates the two $\SU(2)$'s, \ie the $\SU(2)_1$ fields have $\U(1)$ charge $+\frac{1}{2}$, while and the $\SU(2)_2$ ones $-\frac{1}{2}$.

The fields of the 4d $\SU(N)$ D3-brane theory are 6 real scalars combined into 3 complex fields $\Phi_i$, gauge fields $A_\mu$ and for fermions one $\SO(3,1)$ Majorana spinor $\Lambda_A$ and 3 $\SO(3,1)$ Majorana spinors $\Psi_{iA}$, where $A$ is a Majorana spinor index and $i$ is an $\SU(2)$ index.  The bosonic T-spinors are found by diagonalising the kinetic term for the scalars $\Phi_i$ around the fuzzy sphere background
\be
\int \delta\Phi_i^\dagger[(1+J^2)\delta_{ij}-i\epsilon_{ijk}J_k]\delta \Phi_j\;,
\ee
where $\delta\Phi_i=a_i+ib_i$, so only a diagonal part of the $\SO(3)$ rotating $a_i$ and the $\SO(3)$ rotating $b_i$ survives as the $\SU(2)\simeq\SO(3)$ symmetry of the action, together with a $\U(1)_T$.  The (complete set of) eigenvectors of this kinetic operator are given by the vector spherical harmonics $J_i Y_{lm}$ and the spinorial spherical harmonics $\Omega^{\hat{\a}}_{jlm}$. This kinetic operator is diagonalised by defining T-vectors $n_a$ coming from the vector spherical harmonics and T-spinors $\xi^{\hat{\a}}$ coming from the spinor spherical harmonics. Similarly, the kinetic operator for the $\Lambda_A,\Psi_{iA}$ fermions is
\be
\int [i\bar\Psi_i\epsilon_{ijk}J_k\Psi_j+2i\bar\Psi_iJ_i\Lambda-\bar \Psi_i \Psi_i]
\ee
and one expands in the same set of complete eigenvectors of the previous operator. After diagonalising, one defines T-spinors $\zeta^{\hat\a}$ coming from the spinor spherical harmonics, T-scalars $\Lambda$ (the same $\Lambda_A$ from before) and T-vectors $g_a$ coming from the scalar/vector spherical harmonics.

Thus analysing the kinetic operators of the deconstructed theory, one finds that its symmetries are $[\SU(2)_1\times \SU(2)_2]_{diag}\equiv\SU(2)_{\perp}$ and $\U(1)_T\equiv\U(1)$ exactly as in the compactified MN theory and as reflected in the final action, whose twisted part is shown in (\ref{twistDorey}).  However, note that one would initially have been compelled to call $\SU(2)_1$ the  $\SU(2)$ parallel to the sphere directions, and $\SU(2)_2$ the one transverse to them.  The $\U(1)_T$ charge is formally the same as the diagonal $\U(1)$ charge inside the two $\SU(2)$'s.

Note also that in the above construction, all the fields on the classical $S^2$ appeared as limits of functions expanded in the scalar fuzzy spherical harmonics, $Y_{lm}(J_i)$, and the various tensor structures of $S^2$ fields were made manifest by diagonalising their kinetic operators.

{\bf Supersymmetry}. As seen in (\ref{twistDorey}), the kinetic term for the T-scalar/T-vector combination is $\omega^{ab}G_{ab}\Lambda$, and we saw that $G_{\theta\phi}$ decomposes in $ \sum_{l\geq 1, m} \frac{1}{\sqrt{l(l+1)}}\Delta_{S_2}Y_{lm}$, whereas $\Lambda$ decomposes in $Y_{lm}$ for $l\geq 0$. Hence, the minimum, $l=1$ mode of $G_{ab}$ couples with the $l=1$ mode of $\Lambda$, giving a mass term $\sim g_{(l=1)}\Lambda_{(l=1)}$ and as a result all $l\geq 1$ modes for both $\Lambda$ and $g_a$ are massive. On the other hand the fermionic T-scalar $\Lambda_{(l=0)}$ mode, that becomes a fermion in 4d, gets no mass term, so we get N=1 massless fermions in 4d, thus ${\cal N}=1$ supersymmetry.\footnote{We have already discussed how the fermionic T-spinors have no zero eigenvalues on the sphere and thus there are no corresponding massless fermions in 4d.}

Finally, even though twisting is in general needed in order to preserve the 16 supersymmetries along the curved space D-brane \cite{Bershadsky:1995qy},\footnote{Of course, the background can   break some of these 16 supersymmetries, but by the fact that a D-brane is an endpoint of   strings, half the total supersymmetry must be preserved.}  whether or not one gets supersymmetries in the {\em dimensionally reduced theory} is not necessarily known. The only restriction is that after dimensional reduction one can have at most ${\cal N}=1$ supersymmetry, and in the case above we recover indeed ${\cal N}=1$. But in principle one could also end up with ${\cal N}=0$ in the dimensionally reduced theory after twisting, \ie no massless fermions. That is what we will obtain in the ABJM case.

\subsubsection{Compactification vs. deconstruction in massive ABJM}\label{ABJMtwist}

We now come return to the case of the fuzzy $S^2$ in the massive ABJM model, resulting in a D4-brane theory.

\subsubsection*{Compactification} 

From the point of view of the $S^2$ compactification of the D4-brane theory, there is an $\SO(2,1)\times \SO(2)_{34}$ (local) Lorentz invariance, and an $\SO(5)_R$ R-symmetry.  Like in the MN case, there is also a global $\SU(2)_A\times \SU(2)_B \subset\SO(5)_R$, and a $\U(1)_A \subset \SU(2)_A$ subgroup. We define the twisted Lorentz symmetry (T-charge) $Q_T=Q_{34}+Q_A$. As before, twisting means that one embeds the $\U(1)$ spin connection (`gauge field of Lorentz invariance') on $S^2$ into the connection of an R-symmetry subgroup (`transverse' gauge field).
 
The D4-brane fields are the 5 real $\phi^m$'s, the gauge field $A_M$ and a 16-real-component spinor $\Psi$, which is a $\SO(10,1)$ Majorana spinor obeying the condition $\Gamma^{012}\Psi=-\Psi$. We need to decompose the $\phi^m$'s under $\SU(2)_A\times \SU(2)_B$, by extracting a scalar that corresponds to an overall scale, specifically $\tilde{\phi}=\sqrt{\phi^m \phi_m}$. In the deconstructed theory, this will correspond  to the mode giving the `size of the sphere', $\Phi$. The remaining modes, $z^m=\phi^m/\tilde{\phi}$, with $z^mz_m=1$, decompose as $z^{\a\dot{\a}}$ and will correspond  to the transverse scalars $Q^\a_{\dot{\a}}$, transforming under $\SU(2)_A\times \SU(2)_B$.

The fermionic fields $\Psi$ must also be decomposed. Initially one can think of them as Dirac spinors in the $\bf 4$ of $\SO(4,1)$ and the $\bf 4$ of $\SO(5)$. The compactification reduces
\be
\SO(4,1)\to \SO(2,1)\times \U(1)_{34} \qquad \text{and} \qquad {\bf 4}\to {\bf 2}^{\pm\sfrac{1}{2}} \;,
\ee
while
\be
\SO(5)\to \SU(2)_A\times \SU(2)_B \qquad \text{and} \qquad {\bf 4}\to ({\bf 2,1})\oplus  ({\bf 1,2}) \;,
\ee
which we then label through the $\U(1)_A$ charge and $\SU(2)_B\equiv \SU(2)_\perp$ global symmetry as
\be
\SO(5)\to \U(1)_A\times \SU(2)_\perp \qquad \text{and} \qquad {\bf 4}\to  {\bf 1}^{\pm\sfrac{1}{2}} \oplus {\bf 2}^0 \;.
\ee
For bookkeeping, we keep the Lorentz indices downstairs, $\mu = 0,1,2$ and $i = 3,4$, while the global indices upstairs, $\alpha = 1,2$ and $\dot \alpha = \dot 1, \dot 2$. We next separate the fermions that carry an $\SU(2)_A$ index by labelling them as $\lambda_i^\alpha$, while the ones with an $\SU(2)_B$ global index as $\psi_i^{\dot \alpha}$, and suppressing the $\SO(2,1)$ spinor index. If the latter is a real spinor (so that it is \eg a Majorana spinor of $\SO(2,1)$) one ends up with the correct number of degrees of freedom, as the total will add up to 16 real components.

For the gauge fields one has the same decomposition as in the previous subsection: we leave $A_\mu$ as is and define
\be
n_\pm = \frac{1}{\sqrt 2 }(A_3 \pm i A_4)\;.
\ee

We can now summarise the symmetries for the bosonic fields and their associated $\U(1)_T$ charges in the following 
table  
\begin{center}
\begin{tabular}{| l | c c c c | c |}
\hline
 & $\SO(2,1)$ & $\U(1)_{34}$ & $\U(1)_{A}$ & $\SU(2)_\bot$ & $\U(1)_T$ \\
\hline
$A_{\mu}$ & ${\bf 3}$ & ${0 }$ & ${ 0}$ & ${\bf 1 }$ & ${0 }$ \\
$n_{\pm}$ & ${\bf 1}$ & ${\pm 1 }$ & ${ 0 }$ & ${\bf 1}$ & ${\pm 1}$ \\
$\tilde \phi$ & ${\bf 1}$ & ${ 0 }$ & ${ 0}$ & ${\bf 1}$ & ${0}$ \\
$z^{\alpha \dot \alpha}$ & ${\bf 1}$ & ${ 0 }$ & ${\pm \sfrac{1}{2}}$ & ${\bf 2}$ & ${\pm\sfrac{1}{2}}$ \\
\hline 
\end{tabular}
\end{center}

We can similarly summarise the fermions as 
\begin{center}
\begin{tabular}{| l | c c c c | c |}
\hline
 & $\SO(2,1)$ & $\U(1)_{34}$ & $\U(1)_A$ & $\SU(2)_\perp$ & $\U(1)_T$\\
\hline
$\lambda^{\alpha}_i$ &  ${\bf 2}$ & ${\pm \sfrac{1}{2}}$ & 
${ \pm \sfrac{1}{2}}$ & ${\bf 1}$ & ${ 0_2, \pm 1}$ \\
$\psi^{\dot{\alpha}}_i$ & ${\bf 2}$ & ${\pm \sfrac{1}{2}}$ & ${ 0}$ & 
${\bf 2}$ & ${\pm \sfrac{1}{2}}$ \\
 \hline
\end{tabular}
\end{center}
The five-dimensional fields can then be split up according to their T-charge as   
\begin{center}
\begin{tabular}{ l l c l }\label{tab}
\vspace{3pt}
T-scalars: & $Q_T = 0$ && $\tilde \phi$, $A_{\mu}$, $\lambda^{\alpha=2}_{i=1}$,
$\lambda^{\alpha=1}_{i=2}$ \\
\vspace{3pt}
T-spinors: & $Q_T = \pm \sfrac{1}{2}$ && $\psi_i^{\dot\alpha}$, $z^{\alpha\dot\alpha}$ \\
T-vectors: & $Q_T = \pm 1$ && $n_{\pm}$, $\lambda^{\alpha=1}_{i=1}$,
$\lambda_{\alpha=2}^{i=2}$
\end{tabular}
\end{center}
Note that once again we have $\SU(2)_B\equiv\SU(2)_\perp$,  $\U(1)_T$  and the Lorentz $\SO(2,1)$ as the only symmetries of the twisted compactification.

\subsubsection*{Deconstruction}

In deconstructing the sphere from the 3d ABJM theory, one starts with $\SO(2,1)$ Lorentz symmetry and $\U(4)\simeq\SU(4)\times \U(1)_M\simeq\SO(6)_R\times \U(1)_M $ R-symmetry.  The $\U(1)_M$ is a common phase of the $C^I$ scalars, which therefore corresponds  to the M-theory direction that is compactified through identification by the $\mathbb Z_k$ orbifold action. Hence, when going to the IIA theory in order to match with the D4-brane picture by taking $k\to\infty$, the $\U(1)_M$ is broken and one is just left with $\SO(6)_R$. Moreover, as in the MN case, the $\SO(6)_R$ R-symmetry is broken by the choice of the fuzzy sphere vacuum to $\SU(2)_1\times \SU(2)_2\times \U(1)$, with the $\U(1)$ having opposite charges for the two $\SU(2)$'s.

The picture for the mass-deformed ABJM theory around the fuzzy $S^2$ in terms of twisted fields (carrying T-charge) is obtained {\it after} `pulling out' a $\tilde{G}^\alpha$ for all the bifundamental fields, so that one is left with the adjoints of $\U(N)$ or $\U(\bar N)$, as in (\ref{expansi}). Then, the functions on the sphere are actually sections of the appropriate bundle; either ordinary functions, sections of the spinor or the line bundle.  Specifically, anything without an $\alpha$ index is a T-scalar, one $\a$ index means a T-spinor and two $\a$ indices means a T-scalar plus a T-vector in a $( {\bf 1}\oplus {\bf 3})$ decomposition, \ie the $\U(1)_T$ invariance is identified with the $\SO(2)_L\simeq \U(1)_L$ Lorentz invariance of the sphere, described by the index $\a$.  Then $\SU(2)_{2}$ is identified with $\SU(2)_\perp$.

This agrees with what we have already obtained from the bosonic action, since $r^\alpha \sim (r, s^\alpha_\beta)$ gave rise to 2 T-scalars (minus one, which due to the Higgsing becomes the physical $A_\mu$ polarisation) and 2 T-vector degrees of freedom, while $Q^\a_{\dot\a}$ should give rise to 4 bosonic T-spinor degrees of freedom. On the other hand the fermions $\psi_I$ are general (Dirac) 3d spinors, giving 8 complex components, or 8 on-shell degrees of freedom. By extracting a $\tilde{G}^\a$ matrix, one obtains real objects, as was seen for instance in \cite{Nastase:2009ny}, where the complex parallel scalars $r^\a$ decomposed into the real objects $r$ and ${s^\a}_\b$. Similarly, $\chi_\a^{\dot\a}$ in Eq.\;(\ref{expansi}) is real, as are $\tilde{\psi}$ and ${U_\a}^\b$.

The transverse fermions $\chi^\a_{\dot \alpha}$ have 8 real components thus 4 on-shell T-spinor degrees of freedom.  The parallel fermions $\psi_\alpha$ get split into $(\tilde\psi, {U_\alpha}^\beta)$, that is 8 real components, which will give another 2 fermionic on-shell T-scalar degrees of freedom (that we will call $\Lambda$ in the following) and 2 fermionic on-shell T-vector degrees of freedom (that we will call $g_a$ in the following).

In summary, the decomposition under deconstruction matches the decomposition of the last table in  the previous subsection for the twisted compactification as follows: The $\SU(2)_\perp$ invariance matches with $\SU(2)_2$, while the $\U(1)_T$ matches the $\SO(2)_L\simeq\U(1)_{Lorentz}$ symmetry of the $\a$ index. The $\tilde{\phi}$, $A_\mu$, $n_{\pm}$ fields match the ones coming from $r^\a$, that is $\tilde{\phi}$, $A_\mu$, $A_a$, the $z^{\dot\a   \a}$ match $Q_\a^{\dot\a}$, the $\psi_i^\a$ match $\chi_\a^{\dot\a}$, and finally the $\lambda_i^\a$ match the $\psi_\a$ fields, \ie $\Lambda$ and $g_a$.

\subsubsection{Twisting the transverse scalars}

We move on to show how the above assignments of fields can be obtained from our fluctuation action. In \cite{Nastase:2009ny} the transverse scalar kinetic term on the sphere before any rescalings was given by
\be\label{kineticpaper}
2\mu^2 N^2 \int  d^3 x d^2\sigma \sqrt {\hat h}\;
\hat h^{ab}\Big[{(\tilde \nabla_a)^\alpha}_\gamma Q^{\dot\a}_\a{(1+x_k\ts_k)^\gamma}_\b {(\tilde \nabla_b)^\b}_\mu Q^{\mu}
_{\dot\a}\Big]\;,
\ee
where the $Q$'s can be obtained by extracting the Killing spinor $\tilde G^\alpha$ part out of the transverse scalars $q^{\dot\alpha}$ according to
\be
q^{\dot \alpha} =  Q^{\dot\alpha}_\alpha \tilde{G}^\alpha\qquad\text{and}\qquad
q^\dagger_{\dot \alpha} = \tilde{G}^\dagger_\alpha Q_{\dot\alpha}^{ \alpha}\;,
\ee
with
\be
Q^{\dot\alpha}_{\alpha} = \Big(Q^{\dot\alpha}_1\;,\; Q^{\dot\alpha}_2 \Big)\qquad\text{and}\qquad Q_{\dot\alpha}^{\alpha} = \doublet{Q_{\dot\alpha}^1}{ Q_{\dot\alpha}^2} 
\ee
and with the definition of the covariant derivative appearing in the above as
\be\label{covariant}
{(\tilde \nabla_a)^\alpha}_\gamma =\d_a \delta^\a_\gamma +\frac{1}{-2i}\hat h_{ab} K_j^b{(\ts_j)^\alpha}_\gamma\;.
\ee

The expression (\ref{kineticpaper}) does not look like a conventional scalar kinetic term on the sphere for the fields $Q$. However, we will next show how this expression, as well as the associated 3d kinetic term, can be converted into a kinetic term for bosonic {\it spinors} on $S^2$.

We observe that the `covariant derivative' defined\footnote{Here we denote this `covariant derivative' with a 
tilde $\tilde \nabla$, to differentiate it from the proper covariant derivative for spinors on the sphere 
$\nabla$.} in (\ref{covariant}) can be related to the usual covariant derivative for spinors on $S^2$ through (\ref{diracs}) as 
\be\label{Dorey}
{(\tilde \nabla_a)^\alpha}_\gamma =  {\Big[\d_a - \frac{i}{2} S\Gamma_\a S^{-1}\Big]^\alpha}_\gamma
={(S\nabla_a S^{-1})^\alpha}_\gamma - i {(S \Gamma_a S^{-1})^\alpha}_\gamma\;.
\ee
Note that the above is almost the $S$-rotated $\nabla_a$. Then, using (\ref{Dorey}), Eq.\;(\ref{kineticpaper}) becomes
after some algebra
\begin{multline}\label{117}
 - 4\mu^2 N^2\int d^3 x d^2\sigma \sqrt {\hat h} \; \hat h^{ab}\Big[ (Q^{\dot\a} S P_-)_\kappa {(\nabla_a\nabla_b)^\kappa}_\nu(P_- S^{-1} Q_{\dot\a})^{\nu} -  i ( Q^{\dot\a} SP_-)_\kappa {(\nabla_a)^\kappa}_\lambda{(\Gamma_b)^\lambda}_\nu(P_+ S^{-1} Q_{\dot\a})^{\nu}\\
 \qquad+ \frac{1}{\sqrt{\hat h}}\d_a (\sqrt{\hat h})( Q^{\dot\a}S P_-)_\lambda {(\nabla_b)^\lambda}_\nu(P_- S^{-1} Q_{\dot\a})^{\nu}-i  \frac{1}{\sqrt{\hat h}}\d_a (\sqrt{\hat h}) ( Q^{\dot\a}SP_-)_\lambda {(\Gamma_b)^\lambda}_\nu(P_+S^{-1} Q_{\dot\a})^{\nu}\Big]
\end{multline}
up to  total derivatives. Exactly as in the case for the transverse fermions of Section\;\ref{transferm}, it is evident from the above that one could consider the full complex 
Weyl-projected $(P_-S^{-1}Q_{\dot\a})^\a$ spinor as the correct variable. Alternatively, one can reorganise them in terms of (modified) Majorana spinors, by writing
\begin{align}
 (Q^{\dot \alpha} S P_+)_\alpha = &(C^{\dot \alpha}, 0)& (Q^{\dot \alpha} S P_-)_\alpha& = (0, D^{\dot \alpha}) &\\
(P_-S^{-1} Q_{\dot \alpha})^\alpha& = \doublet{0}{-C_{\dot \alpha}}  & (P_+S^{-1} Q_{\dot \alpha})^\alpha & = \doublet{D_{\dot \alpha}}{0}\;,&
\end{align}
where $C^{\dot \alpha}$, $D^{\dot\alpha}$ can be evaluated explicitly, and considering the following combination
\be
\Xi^{\dot \alpha}_\alpha \equiv (Q^{\dot \alpha} S P_+)_\alpha - i (Q^{\dot \alpha} S P_-) = (C^{\dot \alpha}, - i D^{\dot \alpha})
\ee
with
\be
\Xi^\alpha_{\dot \alpha} = \epsilon^{\alpha\beta}\Xi_{\dot \alpha\beta}= \doublet{-i D_{\dot \alpha}}{-C_{\dot \alpha}}   = - i(P_+ S^{-1} Q_{\dot \alpha})^\alpha + (P_- S^{-1} Q_{\dot \alpha})^\alpha
\ee
and the conjugate spinor being 
\be
\bar \Xi_\alpha^{\dot \alpha} \equiv (\Xi^{\dot \alpha\beta})^\dagger {(- \Gamma^3)^\beta}_\alpha =  -i\Big((Q^{\dot \alpha} S P_+)_\beta - i (Q^{\dot \alpha}S P_-)_\beta \Big) {( \Gamma^3)^\beta}_\alpha= -i (C^{\dot \alpha}, iD^{\dot \alpha})\;.
\ee
We now re-write the action in terms of the Majorana spinors $\Xi_{\dot\a}^\a$. Before  performing the substitutions, note that the spin covariant derivative and the $\Gamma_a$'s do not commute
\be
\nabla_a \Gamma^a = \Gamma^a \nabla_a - \frac{i}{2} \Gamma^a \omega^{ij}_a \sigma_{ij}\;.
\ee
Let us first concentrate on the last three terms of (\ref{117}). These give 
\be\label{124}
- 4\mu^2 N^2\int d^3 x d^2\sigma \sqrt {\hat h}\Big[   - D^{\dot \alpha} C_{\dot \alpha} - D^{\dot \alpha} \cot \theta  \d_\theta C_{\dot \alpha} \Big]\;,
\ee
where we have used the identity
\be
\Big(- i\d_\theta +\frac{1}{\sin{\theta}}\d_\phi- \frac{i}{2}\cot{\theta} \Big) D_{\dot \alpha} = -C_{\dot \alpha}\;,
\ee
which can be proved by explicit evaluation. One can easily express $-D^{\dot \alpha} C_{\dot \alpha} = \frac{1}{2}\bar \Xi^{\dot \alpha}_\alpha \Xi_{\dot \alpha}^\alpha$ but we are then left with an extra term. However, note that by calculating the following quantity
\bea
&&-  2 \mu^2 N^2\int d^3 x d^2\sigma \sqrt {\hat h}\Big[ \bar \Xi^{\dot \alpha}_\alpha {(\nabla_a \nabla^a)^\alpha}_\beta \Xi_{\dot \alpha}^\beta\Big]\cr
&=&- 4 \mu^2 N^2\int d^3 x d^2\sigma \sqrt {\hat h}\Big[ - D^{\dot \alpha} {(\nabla_a \nabla^a)^2}_2 C_{\dot \alpha} - D^{\dot \alpha}\cot \theta \d_\theta C_{\dot \alpha} + \frac{1}{2} D^{\dot \alpha}C_{\dot \alpha}\Big]\;,
\eea
up to total derivatives, the first term in the above expression is also the first term of (\ref{117}). Hence
\bea
&&- 4\mu^2 N^2\int d^3 x d^2\sigma \sqrt {\hat h} \; \Big[ (Q^{\dot\a} S P_-)_\kappa {(\nabla_a\nabla^a)^\kappa}_\nu(P_- S^{-1} Q_{\dot\a})^{\nu}\Big]\cr
&=&\mu^2 N^2\int d^3 x d^2\sigma \sqrt {\hat h}\Big[ - 2 \; \bar \Xi^{\dot \alpha}_\alpha {(\nabla_a \nabla^a)^\alpha}_\beta \Xi_{\dot \alpha}^\beta - 4  D^{\dot \alpha}\cot \theta \d_\theta C_{\dot\alpha} + 2 D^{\dot \alpha}C_{\dot \alpha}\Big]
\eea
and the middle term will cancel the similar contribution from  (\ref{124}). Finally, with the definition of the Dirac operator on a sphere of unit radius given by\footnote{In the Dirac operator  $\hat\nabla_{S^2}$ the indices are raised and lowered with the unit sphere metric $\hat h_{ab}$.}
\be
-i \hat \nabla_{S^2} = -i \Gamma^a \nabla_a
\ee
 and since
\be
\nabla_a \nabla^a \otimes \one_{2 } = \Gamma^a \nabla_a \Gamma^b \nabla_b\;,
\ee
the resulting expression for the transverse scalar kinetic term on the sphere is
\be\label{transca}
N^2\int d^3 x d^2\sigma \sqrt {\hat h}\Big[   \frac{1}{2} \bar \Xi^{\dot \alpha}_\alpha {((- i2 \mu\hat\nabla_{S^2})^2)^\alpha}_\beta \Xi_{\dot \alpha}^\beta  - 3 \mu^2\bar \Xi^{\dot \alpha}_\alpha  \Xi_{\dot \alpha}^\alpha \Big]\;.
\ee
To that we need to add the 2+1d kinetic term, which can be easily evaluated to be
\bea
&&-\frac{N^2}{2}\int  d^3 x d^2\sigma \sqrt {\hat h}\; \Big[ \d_\mu Q^{\dot \alpha}_\alpha (\delta^\alpha_\beta + x_i {(\tilde \sigma_i)^\alpha}_\beta)\d^\mu Q^\beta_{\dot \alpha}\Big]\cr
&=&N^2\int  d^3 x d^2\sigma \sqrt {\hat h}\; \Big[ -\frac{1}{2} \d_\mu \bar \Xi_\alpha^{\dot \alpha}\d^\mu \Xi^\alpha_{\dot \alpha}\Big]\;.
\eea

The final answer for the transverse scalars in terms of T-spinors on the sphere is
\be\label{finaltransverse}
N^2\int d^3 x d^2\sigma \sqrt {\hat h}\Big[   \frac{1}{2} \bar \Xi^{\dot \alpha} (- i2 \mu\hat\nabla_{S^2})^2 \Xi_{\dot \alpha}  -\frac{1}{2} \d_\mu \bar \Xi^{\dot \alpha}\d^\mu \Xi_{\dot \alpha} - 3 \mu^2\bar \Xi^{\dot \alpha}  \Xi_{\dot \alpha} \Big]\;.
\ee
One might be worried about the fact that the kinetic term for the T-spinor on the sphere is quadratic in the Dirac operator. However, as also argued in \cite{Andrews:2006aw}, this makes sense as the kinetic term for a boson is quadratic in derivatives and it is not possible to lose a derivative from our above redefinition of fields.

\subsubsection{Twisting the parallel fermions}

We finally decompose the parallel fermions $\psi_\a$ into bosonic quantities, after extracting an explicit $\tilde G^\alpha$, due to the bifundamental nature of $\psi_\a$. This is implemented using the form
\be
\psi_\a= \tilde \psi \tilde{G}_\a + {U_\alpha}^\beta \tilde{G}_\beta ,
\ee
with $\tilde\psi$ and ${U_\a}^\b$ decomposing in terms of fuzzy spherical harmonics $Y_{lm}(J_i)$. 
The expansion of the Hermitian conjugate is
\be
\psi^{\dagger \a}=  \tilde{G}^{\dagger\a}\bar{\tilde \psi} + 
\tilde{G}^{\dagger\beta} \bar U_\beta^{\phantom{\beta}\alpha}, 
\ee
however note that in 2+1d the fermions are actually Majorana, so $\bar\psi=\psi^T C$, \ie is related by simply raising/lowering  with $\epsilon^{\sigma\tau}$ in the suppressed 3d spinor indices.

We next define
\be
{U_\a}^\b=\frac{1}{2}U_i{(\ts_i)_\a}^\b\;,\qquad
{\bar U_\a\,}^\b=\frac{1}{2} U_i {(\ts_i)_\a}^\b
\ee
and then split the above into a transverse and two tangential components on the sphere as
\be
U_i=K_i^a g_a+\hat\psi x_i\;,\qquad
\bar U_i=K_i^a\bar g_a+\bar{\hat\psi} x_i\;.
\ee
Having done that, we are now ready to express the parallel fermionic components of (\ref{abjmfermions}) in terms of the fields $g_a$ and $\Lambda$. 

\subsubsection*{Mass terms}

For the mass deformation we find\footnote{Note that the $g_a$ and $\hat \psi$ fields obey Fermi statistics, and one can easily prove that \eg $\bar U_i \tilde\psi=\bar{\tilde\psi} U_i$.} 
\bea
-i\mu\Tr[\psi^{\dagger\alpha}\psi_\alpha] &=& -i\mu
\Tr[-(N-1) \bar{\tilde \psi}\tilde\psi + {J_\alpha}^\beta {\bar U_\beta\,}^
\alpha \tilde \psi + {J_\b}^\a\bar{\tilde \psi} {U_\a}^\b  +{J_\gamma}^\b{\bar U_\b\,}^\a {U_\a}^\g]\cr
&\rightarrow& i\mu N^2 \int d^2\sigma \sqrt{\hat h }[ \bar{\tilde \psi}\tilde \psi 
-  \bar{\tilde \psi} \hat\psi + \frac{1}{4}\bar{\hat \psi}\hat \psi+\frac{1}{4} \hat h^{ab}  \bar g_a g_b]\;.
\eea

\subsubsection*{3d kinetic terms}

Similarly to the above one has
\bea
- \Tr[\psi^{\dagger\alpha}\slashed{\partial}
\psi_\alpha] &=& - \Tr[-(N-1) \bar{\tilde \psi}\slashed{\partial}\tilde\psi + 
{J_\alpha}^\beta {\bar U_\beta\,}^\alpha \slashed{\partial}\tilde \psi + {J_\b}^\a\bar{\tilde \psi} 
\slashed{\partial}{U_\a}^\b  +{J_\gamma}^\b{\bar U_\b\,}^\a \slashed{\partial}{U_\a}^\g]\cr
&\rightarrow&  N^2 \int d^2\sigma \sqrt{\hat h }[ \bar{\tilde \psi} \slashed{\d}\tilde \psi -\bar{ \hat \psi} \slashed{\d} \tilde \psi+ \frac{1}{4}\bar{\hat \psi}\slashed{\d} \hat \psi+\frac{1}{4} \hat h^{ab}  \bar g_a \slashed{\d} g_b]\;.
\eea
 
\subsubsection*{$CC\psi\psi$ terms: $\tilde \psi^2$-terms}

We can use as a starting point Eq.\;(\ref{28}) and expand the fields accordingly
\bea\label{41}
&&-2\mu N  \int d^3 x d^2 \sigma  \sqrt{\hat h}\; \Big[\psi_\a({(S\Gamma^a S^{-1})^\a}_\b \d_a\psi^{\dagger \b})- \frac{i}{2}{(S \Gamma_3 S^{-1})^\a}_\b \psi_\a\psi^{\dagger \beta}\Big]\cr
&&=2\mu N  \int d^3 x d^2 \sigma  \sqrt{\hat h}\; \Big[ {(S\Gamma^a S^{-1})^\a}_\b  G_\a\d_a(G^{\dagger \b}) 
\bar{\tilde \psi}\tilde{\psi} \cr
&&\qquad\qquad\qquad\qquad\qquad+ {J_\a}^\b {(S\Gamma^a S^{-1})^\a}_\b   (\d_a\bar{\tilde \psi})\tilde{\psi}
- \frac{i}{2} {J_\a}^{\beta}{(S \Gamma_3 S^{-1})^\a}_\b \bar{\tilde\psi}\tilde\psi\Big]\;.
\eea
Using ${J_\a}^\b\rightarrow \frac{N}{2}({x_k\ts_k-1)^\b}_\a$
and substituting the relations (\ref{SKx}) and the traces (\ref{traces})
in the final line of (\ref{41}), we get a simple mass term 
\be
3i\mu N^2  \int d^3 x d^2 \sigma  \sqrt{\hat h}\;  \bar{\tilde\psi}\tilde\psi \;.
\ee

\subsubsection*{$CC\psi\psi$ terms: $U^2$-terms}

For the $U^2$-type terms we have
\bea\label{542nd}
&&2\mu N  \int d^3 x d^2 \sigma  \sqrt{\hat h}\; \Big[{(S\Gamma^a S^{-1})^\a}_\b (\d_a\psi^{\dagger \b})\psi_\a- \frac{i}{2}{(S \Gamma_3 S^{-1})^\a}_\b \psi^{\dagger \beta}\psi_\a\Big]\cr
&&=2\mu N  \int d^3 x d^2 \sigma  \sqrt{\hat h}\; \Big[{(S\Gamma^a S^{-1})^\a}_\b \d_a(G^{\dagger \gamma} {\bar 
U_\gamma\,}^\beta)({U_\alpha}^\mu G_\mu) - \frac{i}{2} {(S\Gamma_3 S^{-1})^\a}_\b G^{\dagger \gamma} {\bar
U_\gamma\, }^\beta {U_\alpha}^\mu G_\mu\Big]\;,\nn\\
\eea
which  can be split into a term involving a derivative and one without a derivative. For the first
we obtain
\bea\label{T1T2}
&&-\frac{\mu N^2}{4}\int d^3 x d^2 \sigma  \sqrt{\hat h}\; \Big[\frac{i}{2}(\delta_{li}-x_lx_i)\Tr[(x_m\ts_m-1)
\ts_k\ts_l\ts_j\ts_i]\bar U_j U_k\cr
&&\qquad\qquad\qquad\qquad+K_l^a\Tr[(x_m\ts_m-1)\ts_j\ts_l\ts_i](\d_a\bar U_i) U_j\Big]\;,
\eea
where we have used the identities on the sphere.
Using once again the identities on the sphere and the traces (\ref{traces}), we obtain
\be
\mu N^2  \int d^3 x d^2 \sigma  \sqrt{\hat h}\; \Big[  -\frac{iN}{2}\hat\omega^{ab}\bar G_{ab}\hat\psi 
+\frac{i}{2}\bar{\hat\psi}\hat\psi\Big]\;,
\ee
where $G_{ab}= \d_a g_b - \d_bg_a$ and $\bar g_a g_b=\bar g_b g_a$. For the second term in (\ref{542nd}) we obtain similarly
\be
 \mu N^2  \int d^3 x d^2 \sigma  \sqrt{\hat h}\; \Big[-\frac{i}{4}\hat h^{ab}\bar g_a g_b +\frac{i}{4} \bar{\hat\psi}
\hat \psi\Big]\;.
\ee

In total for the $U^2$-type terms we get the simple expression
\be
i \mu N^2  \int d^3 x d^2 \sigma  \sqrt{\hat h}\; \Big[ -\frac{1}{2}\hat\omega^{ab}\bar G_{ab}\hat\psi +
\frac{3}{4}\bar{\hat\psi} \hat \psi
- \frac{1}{4}\hat h^{ab} g_a g_b \Big]\;.
\ee
Note that the mass term $\bar g^a g_a$ here precisely cancels the mass term coming from $\psi^{\dagger\alpha} \psi_\alpha$.

\subsubsection*{$CC\psi\psi$ terms: $U$-$\tilde \psi$-terms}

For the $U$-$\tilde \psi$-type terms we have 
\bea\label{54}
&&2\mu N  \int d^3 x d^2 \sigma  \sqrt{\hat h}\; \Big[{(S\Gamma^a S^{-1})^\a}_\b (\d_a\psi^{\dagger \b})\psi_\a- \frac{i}{2}{(S \Gamma_3 S^{-1})^\a}_\b \psi^{\dagger \beta}\psi_\a\Big]\cr
&&=2\mu N  \int d^3 x d^2 \sigma  \sqrt{\hat h}\; \Big[{(S\Gamma^a S^{-1})^\a}_\b [\d_a(G^{\dagger \beta}
\bar{\tilde \psi} ) {U_\alpha}^\mu G_\mu +\d_a( G^{\dagger \gamma} {\bar{U}_\gamma\,}^\beta)\tilde \psi G_\alpha] \cr
&&\qquad\qquad\qquad\qquad\qquad - \frac{i}{2} {(S\Gamma_3 S^{-1})^\a}_\b[ G^{\dagger \gamma} {\bar U_\gamma\, }
^\beta \tilde \psi G_\alpha +G^{\dagger \beta}\bar{\tilde \psi} {U_\alpha}^\mu G_\mu]\Big]\;.
\eea
Through identities and manipulations all of which  we have already seen, we get for the first term 
\bea
&=&\mu N^2\int d^3 x d^2 \sigma  \sqrt{\hat h}\; \Big[ (\d^a \bar g_a) \tilde \psi+\frac{1}{\sqrt{ \hat h}}(\d^a\sqrt{\hat h})\bar g_a \tilde{\psi}+(\d^a\bar{\tilde\psi})g_a - 2i  \bar{\tilde \psi} \hat \psi+i \hat\omega^{ab}\bar G_{ab} \tilde \psi\Big]\cr
&=&\mu N^2\int d^3 x d^2 \sigma  \sqrt{\hat h}\; \Big[  - 2i  \bar{\tilde \psi} \hat \psi+i\hat\omega^{ab}\bar G_{ab}
\tilde{\psi}\Big]\;.
\eea
Since the fermion scalar products are symmetric under exchange, the first three terms in the first line above have cancelled. For the second term in (\ref{54}) we get
\be
-i \mu N^2\int d^3 x d^2 \sigma  \sqrt{\hat h}\;\Big[\bar{\tilde \psi} \hat \psi\Big]  \;.
\ee
The  final result for the $U$-$\tilde \psi$-terms is simply
\be
\mu i N^2\int d^3 x d^2 \sigma  \sqrt{\hat h}\; \Big[  -3\bar{ \tilde \psi} \hat \psi
+\hat\omega^{ab}\bar G_{ab}\tilde{\psi}\Big]\;.
\ee

\subsubsection*{Total result for the parallel fermions}
Collecting all the terms the total result from expanding the parallel fermions is
\bea
 &\to&  N^2 \int d^2\sigma \sqrt{\hat h }[ (\bar{\tilde \psi} - \frac{1}{2}\hat{\hat \psi}) \slashed{\d}({\tilde \psi} - \frac{1}{2}\hat \psi)+\frac{1}{4} \hat h^{ab}  \bar g_a \slashed{\d} g_b +  i\mu\hat\omega^{ab}\bar G_{ab}
(\tilde{\psi}-\frac{1}{2}\hat \psi) + 4i\mu(\bar{\tilde\psi}-\frac{1}{2}\bar{\hat\psi})(\tilde{\psi}-\frac{1}{2}\hat{\psi})]
\label{parferm}\cr
&=& N^2 \int d^2\sigma \sqrt{\hat h }[ \frac{1}{4}\bar \Lambda \slashed{\d}\Lambda +\frac{1}{4}  \bar g_a \slashed{\d} g^a + \frac{ i\mu}{2}\hat\omega^{ab}\bar G_{ab}
\Lambda + i\mu \bar \Lambda \Lambda]\;,
\eea
where $\Lambda \equiv 2 ({\tilde \psi} - \frac{1}{2}\hat \psi)$.

\subsubsection*{Disappearance of $\tilde{\psi}+\frac{1}{2}\hat\psi$ parallel fermion mode}

The above result shows that the \emph{a priori} independent quantities $\tilde{\psi}$ and $\hat\psi$ combine into a single mode, in much the same way as it was observed in \cite{Nastase:2009ny} for the scalar $\Phi=2r+\phi$ mode. In that case, the disappearance of the $2r-\phi$ mode was due to the Higgs mechanism with the degree of freedom being eaten by the gauge field, which then became dynamical (Yang-Mills). For the fermions however, half the components are solved by the Dirac equation in terms of the other half, whereas for scalars no component is lost on-shell, and for a 2d gauge field all components are lost on-shell.  For that reason, the twisting is expected to result in losing some components in the classical limit.

In terms of algebra, the explanation is the same as for the case involving the scalars. Start with 
\bea
\psi_\alpha &= &\tilde \psi \tilde{G}_\alpha + {U^\alpha}_\beta \tilde{G}^\beta\cr
&\rightarrow& 
\tilde \psi \tilde{G}_\alpha - \epsilon_{\alpha\beta}\frac{{(\ts_i)^\beta}_\gamma}{2} (K_i^a g_a 
+\frac{J_i}{N} \hat \psi) \tilde{G}^\gamma
\eea
and using 
\be
\frac{{(\ts_i)^\beta}_\gamma}{2N}J_i \tilde{G}^\gamma = {J^\beta}_\gamma \tilde{G}^\gamma - (N-1)
\tilde{G}^\gamma = \frac{N+1}{2}\tilde{G}^\gamma \;,
\ee
one gets at leading $N$
\bea
\psi_\alpha &\rightarrow& \tilde \psi \tilde{G}_\alpha - \epsilon_{\alpha\beta}\frac{{(\ts_i)^\beta}_\gamma}{2} 
K_i^a g_a - \frac{1}{2}\epsilon_{\alpha\gamma} \hat \psi \tilde{G}^\gamma\cr
&=& (\tilde \psi - \frac{1}{2}\hat \psi) \tilde{G}_\alpha + \frac{{(\ts_i)_\alpha}^\gamma}{2} K_i^a g_a 
\tilde{G}_\gamma\cr
&\equiv& \frac{1}{2}\Lambda \tilde{G}_\alpha + \frac{1}{2} K_i^a g_a \tilde{G}_\gamma 
{(\sigma_i)^\gamma}_\alpha\;.
\eea

\subsubsection{Final action}

Collecting all the terms, the result  for the twisted fields is
\begin{multline}\label{finalaction}
S=\int d^3x d^2\sigma \;\sqrt{ \hat h} 
 \Big[-\left(\frac{k}{2\pi}\right)^2\frac{ 1}{8 f^2} F_{\mu\nu}
 F^{\mu\nu} -\frac{N^2 \mu^2}{2}F_{ab}F^{ab} -\frac{N^2}{4}\partial
_\mu A^a \partial^\mu A_a  +Nf \d_\mu A_a\d^a A_\mu\\
-f^2 \d^a A_\mu \d_a A^\mu  + N^2 \mu^2 F_{ab}\hat\omega^{ab}\Phi  - \frac{N^2}{4} \partial_\mu  
\Phi \partial^\mu \Phi- N^2 \mu^2 \d_a \Phi \d^a \Phi- N^2 \mu^2 \Phi^2  \\
+N^2\Big(\Big(\frac{1}{2}\bar \Upsilon^{\dot \alpha} D_5  \Upsilon_{\dot\alpha} + i \mu \bar \Upsilon^{\dot\alpha} 
\Upsilon_{\dot{\a}}+h.c.\Big) 
+\frac{1}{2} \bar \Xi^{\dot \alpha} (- i2 \mu\hat\nabla_{S^2})^2 \Xi_{\dot \alpha}  -\frac{1}{2} 
\d_\mu \bar \Xi^{\dot \alpha}\d^\mu \Xi_{\dot \alpha} -3 \mu^2\bar \Xi^{\dot \alpha}  \Xi_{\dot \alpha}\Big)\\
+N^2\Big(\frac{1}{4}\bar \Lambda \slashed{\d}\Lambda +\frac{1}{4}  \bar g_a \slashed{\d} g^a + 
\frac{i\mu}{2}\hat\omega^{ab}\bar G_{ab}\Lambda + i\mu \bar \Lambda \Lambda\Big)\Big]\;.
\end{multline}
After the rescalings in (\ref{rescabos}) and (\ref{rescaferm}), as well as
\be
\Lambda \to \Lambda \frac{4\pi l_s}{N \mu}\;,\qquad g_a \to g_a \frac{4\pi l_s}{N}\;,\qquad \Xi_\alpha^{\dot\alpha} \to \Xi_\alpha^{\dot\alpha} \frac{4\pi l_s}{N\mu}\;,
\ee
the final action is 
\bea
S_{phys} &=& \frac{1}{g_{YM}^2}\int d^3 x d^2 \s \sqrt { h }\; 
 \Big[-\frac{1}{4}  F_{AB}  F^{AB}-\frac{1}{2} \partial_A 
\Phi \partial^A\Phi - \frac{\mu^2}{2} \Phi^2 
+ \frac{\mu}{2}\; \omega^{ab} F_{ab}\Phi\cr
&&+\Big(\frac{1}{2}\bar \Upsilon^{\dot \alpha} D_5  \Upsilon_{\dot\alpha} + \frac{i}{2} \mu \bar 
\Upsilon^{\dot\alpha} \Upsilon_{\dot{\a}}+h.c.\Big)
+\frac{1}{4} \bar \Xi^{\dot \alpha} (- \frac{2i}{\mu} \nabla_{S^2})^2 \Xi_{\dot \alpha}  -
\d_\mu \bar \Xi^{\dot \alpha}\d^\mu \Xi_{\dot \alpha} -  \frac{3}{2} \mu^2\bar \Xi^{\dot \alpha}  \Xi_{\dot \alpha}\cr
&&+\frac{1}{4}\bar \Lambda \slashed{\d}\Lambda +\frac{1}{4}  \bar g_a \slashed{\d} g^a + 
\frac{i}{4}\omega^{ab}\bar G_{ab}\Lambda + \frac{i}{2}\mu \bar \Lambda \Lambda
\Big]\;.
\eea
This is the twisted action of a D4-brane on $S^2$. Comparing with the twisted action of \cite{Andrews:2006aw}, we see the same kinetic terms appearing for the twisted fields, which is very encouraging. The only difference is in the appearance of the mass terms, which reflect the nontriviality of the background that the D4-brane is probing. This would imply that by dimensionally reducing the theory on the $S^2$ one  has ${\cal   N}=0$ (no supersymmetry), since we still have massless 3d bosons ($A_\mu$) but no massless 3d fermions any more.

But is not so unexpected: In \cite{Andrews:2006aw}, the supersymmetry stayed the same ${\cal   N}=1$ throughout. Starting with a $\SU(N)$ action in 4d with ${\cal N}=1$, and in the classical limit for the fuzzy sphere background, one obtained a twisted 5-brane action that dimensionally reduces back to the same ${\cal N}=1$.  Here by contrast, we start with an ${\cal   N}=6$-invariant action, fuzzy sphere background and fluctuations.  By compactifying the resulting D4-brane action for the fluctuations, we could have at most obtained ${\cal N}=1$ back in 3d, but in any case not ${\cal N}=6$.

\subsection{Fuzzy sphere vs. fuzzy funnel and gravity dual picture}

In this subsection we come back to the issue of the fuzzy funnel in the undeformed ABJM theory. As was shown in \cite{Nastase:2009ny}, the result for the unrescaled finite-$N$ bosonic action is the same for both the fuzzy sphere GRVV ground state and the fuzzy funnel BPS solution of ABJM, with the replacement of $\mu\to \frac{1}{2s}$. Roughly speaking, the reason is that the derivative $\d_s$ played the role of $\mu$ when acting on the funnel profile $Z^\a\propto \frac{1}{\sqrt{s}}$, since $\d_s [\frac{1}{\sqrt{s}}]= [\frac{1}{2s}]\frac{1}{\sqrt{s}}$.  All terms proportional to $\mu$ in the massive case were reproduced for the funnel with this substitution.

In the case of the fermionic finite-$N$ fluctuation action however, the only terms explicitly proportional to $\mu$ are the mass terms from the original GRVV action, and these cannot be reproduced by $\d_s$ acting on $Z^\alpha$ any more, as can be seen simply using dimensional arguments.\footnote{One would need at least a term proportional to $ \bar \psi \d_s C^I \psi$, which besides not being invariant, does not have the right dimension.} Hence, the full unrescaled finite-$N$ fluctuation action around the fuzzy funnel is different from that corresponding to expanding around the GRVV fuzzy sphere vacuum, and in particular is not supersymmetric.

The only alternative possibility to get the `mass' terms would be from the kinetic term $\bar \psi \d_s\psi$, if the fermion $\psi$ were also proportional to $\frac{1}{\sqrt{s}}$. That however means that one would have to accordingly rescale the fermionic fluctuation $\psi$. In fact, one needs to rescale even the bosonic fields in order to get to the standard form of the classical action, but that is also problematic: the derivative $\d_s$ can act on all fields rescaled by $\mu\sim \frac{1}{2s}$-dependent terms in Eq.\;(6.5) of \cite{Nastase:2009ny}. The rescaled action then looks complicated and incomplete.

Nevertheless, let us pause and ask what one would expect to recover: By comparison with the fuzzy sphere action, we want to obtain the action for a D4-brane on $R^{2,1}\times S^2$ in the classical limit, perhaps with extra field configurations turned on in its worldvolume, giving a D2-brane charge. In fact, based on the supersymmetry analysis at the beginning of this section, one expects half the supersymmetry of the D4-brane action, with the system corresponding to a D2-brane ending on a D4 (such that $\Gamma_s\epsilon=\Gamma_{3456}\epsilon$). This would imply that our fluctuation action is missing both the D2-brane charge on the D4 worldvolume as well as the D2-D4 open string degrees of freedom.  Only once these are taken into account, with a correct analysis of the modes along the $s$ worldvolume direction, should one expect to find the correct brane action preserving $\frac{1}{2}$ the supersymmetry of the background. As it is, we can at most deduce that at $s\rightarrow\infty$, when $\mu = \frac{1}{2s}\rightarrow 0$ (but is still large enough so that the fuzzy sphere of radius $R\propto \sqrt{\mu}$ can be considered classical) the fuzzy funnel action coincides with the fuzzy sphere one. In that case, one is far away from the source of D2-brane charge, and thus only the D4-brane action remains. As the full picture for the fuzzy funnel case does not extend straightforwardly from the analysis performed in this paper, we will leave further investigation as an open question for the future.

We conclude this section by providing a spacetime picture for the D4-brane on the fuzzy sphere, to further justify why one naturally recovers such a D4-brane action on $S^2$ in the classical limit. It was argued for the $\cA_4$-theory case in \cite{Gomis:2008cv} that the fuzzy sphere ground state for the massive BLG model in M-theory corresponds to a giant graviton D3-brane in type IIB, wrapping an $S^3$ inside the maximally supersymmetric pp-wave. With the correct $N$-membrane picture being captured by ABJM, one can apply similar arguments for the mass-deformed case of GRVV, where however the classical limit of the fuzzy sphere ground state contains one subtlety: The massive deformation of ABJM still corresponds in IIB to considering the maximally supersymmetric pp-wave background, but now one also has a $\mathbb Z_k$ orbifolding. Moreover, as was also the case in \cite{Nastase:2009ny}, the classical large-$N$ limit together with the condition for small fluctuations forces us to additionally take $k\rightarrow\infty$.

By defining the 4+4 coordinates transverse to the IIB pp-wave as $Z^I=(Z^\a,Z^{\dot{\a}})$ with 
\bea
Z^1=X^1+iX^2\;,&\qquad& Z^2=X^3+iX^4\cr
Z^{\dot 1}=X^5+iX^6\;,&\qquad& Z^{\dot 2}=X^7+iX^8
\eea
(and with $X^\pm=X^0\pm X^9$), then the giant graviton D3-brane wraps a sphere of radius
\be
|Z^1|^2+|Z^2|^2=R^2\;.
\ee
If one then performs a T-duality along one of the $Z^{\dot \a}$ coordinates to type IIA string theory and  lifts up to M-theory on a coordinate $X^{10}$, the configuration becomes an M5-brane wrapping the $S^3$ and also extending in $X^0,X^9,X^{10}$.  Going from BLG to ABJM corresponds to increasing the number of branes to $N$, while also dividing the target space by $\mathbb Z_k$ acting by $Z^i\rightarrow e^{2\pi i/k}Z^i$, which shrinks the $S^1$ Hopf fibre of the $S^3$ fibration over $S^2$ $k$ times. In the {\em classical large-$N$, large-$k$ limit} one must then reduce M-theory to type IIA on the shrunk Hopf fibre coordinate as in \cite{Nastase:2009ny}, instead of  $X^{10}$, to  obtain a D4-brane wrapping a classical $S^2\simeq S^3/S^1$ and also extending in the coordinates $X^0,X^9,X^{10}$.

\section{Conclusions and discussion}\label{discussion}

In this paper we have continued the analysis of the fuzzy sphere/funnel solution for the massive/pure ABJM model initiated in \cite{Nastase:2009ny}. In the latter it was shown that the solutions of \cite{Gomis:2008vc} involved fuzzy 2-sphere configurations, instead of the anticipated 3-spheres, although formulated in terms of bifundamental rather than the usual adjoint matter fields. In this work, we have explicitly expressed these configurations in a way that is completely equivalent to the usual $\SU(2)$ construction. The representations of the GRVV algebra (\ref{algebra}) in terms of the bifundamental generators $\tilde{G}^\a$ are equivalent to the ones of the $\SU(2)$ algebra in terms of the adjoint $J_i$. Moreover, since $(J_i,\tilde G^\a)$ can be packaged neatly in a supermatrix form to give what is known as the `fuzzy supersphere', we additionally obtained the statement that the latter is equivalent to the conventional bosonic fuzzy sphere. In the classical limit, $\tilde{g}^\a=\frac{1}{\sqrt{N}}\tilde{G}^\a$ become Weyl-projected Killing spinors of $S^2$ (up to a phase that cannot be explicitly determined), thus suggesting that the  $\tilde{G}^\a$ can be thought of as fuzzy Killing spinors on the fuzzy $S^2$.  We also presented generalisations of these statements to the $S^4$ and $S^8$ cases (corresponding to the second and third Hopf maps), as well as to $\cp^3$ (corresponding to the embedding of the first Hopf map into the $\cp^3$ Hopf map).

We then obtained the full supersymmetric action for small fluctuations around a D4-brane on $R^{2,1}\times S^2$, starting from the classical (large-$N$) limit of the mass-deformed ABJM model around the fuzzy sphere solutions of \cite{Gomis:2008vc}. This was done by completing the bosonic part of the fluctuation action, treated in \cite{Nastase:2009ny}, through the evaluation of the fermionic piece. The latter presented some interesting new features compared to the bosonic case. In particular, it raises the question about how the spinorial spherical harmonic expansion of spinors appears on the fuzzy sphere. In the usual (adjoint) `deconstruction' approach of \cite{Andrews:2006aw} for the fuzzy $S^2$, one first obtains scalar functions expanded in the scalar fuzzy spherical harmonics $Y_{lm}(J_i)$, with the spinor or tensor structure appearing by diagonalising the kinetic operators in the classical limit. In our present (bifundamental) construction, a natural guess would have been to expand in terms of $Y_{lm}(x_i)\tilde{g}^\a$, given the relation (\ref{killspsphhar}) between Killing spinors and spinorial spherical harmonics. However, there are subtle points to this argument. If one keeps the bosonic/fermionic structure dictated by the 3d part of the action, then the combination $Y_{lm}(x_i)\tilde{g}^\a$ does indeed appear not only for fermions but also for bosons, and in particular for the set of transverse scalars $q^{\dot\a}$. Moreover, even though we did obtain the usual maximally supersymmetric D4-brane action, the transition from finite to infinite $N$ becomes harder to understand, \eg the supersymmetry of the final action cannot be straightforwardly obtained from the finite-$N$ version.

The issue of supersymmetry for D-brane worldvolumes on curved spaces is intimately linked with the issue of twisting.  We have reviewed why twisting is necessary when compactifying a D-brane theory, and how it can appear when `deconstructing' the theory, by revisiting the closely related case of \cite{Andrews:2006aw}. We then applied a similar logic to our problem of interest and found a twisted supersymmetric D4-brane action, for which supersymmetry is easier to understand. A comparison with \cite{Andrews:2006aw} yields various similarities but also significant differences. In particular, in the mass-deformed ABJM case the dimensionally-reduced, `deconstructed' theory we naturally obtain preserves no supersymmetry, which is perhaps unexpected though not inconsistent.  An interesting consequence of this analysis is that the fuzzy Killing spinor $\tilde{G}^\a$ allows a unified presentation of twisted and untwisted fields, with the process of twisting reducing to adding or subtracting a $\tilde{G}^\a$.

We should comment on the fact that one could never obtain a classical M5-brane action for the M2-M5 system described by the large-$N$ fuzzy sphere background in this way. As explained in the introduction, in perturbation theory we are forced to take large $k$ together with large $N$, and hence the $\mathbb Z_k$ reduction turns the $S^3$ into an $S^2$ by modding out the $S^1$ fibre of the Hopf fibration. But should one expect to find a classical M5-brane action in some limit, perhaps by computing the full D4-brane action, not just in the approximation of dealing with quadratic fluctuations?

The action of multiple M5-branes is expected to be conformal and hence to have no coupling constant associated with it. As a result, in a perturbative expansion (for small fluctuations) of any kind, one should not expect to see the appearance of an M5.  Moreover, the D4-brane coupling is given by
\be
 g_{YM}^2=g_s l_s = R_{11}=\sqrt{\frac{N}{k}\mu l_p^3}\;, 
\ee
 therefore in the D4-brane perturbation theory that we uncovered, \ie the quadratic action for which the  ABJM coupling $\lambda =\frac{N}{k}$ is kept fixed and small, one is always in the type IIA regime. By definition, the M5 appears at infinite D4-brane coupling, \ie  when $N$ is infinite, if $k$ is of order 1. In that case however, one would have to have knowledge of the full quantum D4-brane action. It follows that it is impossible to explicitly see the M5-brane appearing in the classical limit of the fuzzy sphere ground state. The M-brane dynamics would only emerge in the strong coupling limit of the theory.

It is important to note that, to the best of our knowledge, there is still a puzzle relating to a discrepancy in the counting of vacua between the mass-deformed gauge theory of \cite{Gomis:2008vc} and the dual geometries of \cite{Bena:2004jw,Lin:2004nb}. Resolving this issue, as well as completely understanding the space of solutions of the theory, is of significant interest for the following reason: Solutions of the GRVV algebra in terms of same-size reducible representations should correspond to coincident multi-D4-brane configurations wrapping the same $S^2$, as argued in Appendix C of \cite{Nastase:2009ny}. It should be straightforward but essential to show that at the level of the fluctuation action. Then in the strong coupling ($k=1$) limit one would recover a configuration of multiple, parallel M5-branes of M-theory in flat space, albeit with zero net M5-brane charge, in the same way that in the same limit of ABJM one recovers multiple M2-branes.

\section*{Acknowledgements}
\noindent 
We would like to thank Sanjaye Ramgoolam for collaboration at the initial stages of this work; Aki Hashimoto for many useful discussions, and also Manavendra Mahato, Shiraz Minwalla and Takeshi Morita for further discussions and comments. The research of HN is partially supported by MEXT's program ``Promotion of Environmental Improvement for Independence of Young Researchers'' under the Special Coordination Funds for Promoting Science and Technology, and also partially by MEXT KAKENHI grant nr.  20740128.

\begin{appendix}

\section{Identities on the sphere}\label{AppA}
In this appendix we repeat some useful identities presented in \cite{Nastase:2009ny}. In obtaining the action for fluctuations on the (unit) classical sphere one needs to make use of a set of Killing vectors $K_i^a$. The explicit formulae for the latter are given by
\bea
K_1^\theta = -\sin{\phi} && K_1^\phi = -\cot{\theta}\cos{\phi}\cr
K_2^\theta = \cos \phi~~~  && K_2^\phi = -\cot{\theta}\sin{\phi}\cr
K_3^\theta = 0~ ~~~~~~~&& K_3^\phi = 1\;,
\eea
 as in \cite{Papageorgakis:2005xr}. The relations between Cartesian and spherical
 coordinates is
\bea
x_1 &=& \sin{\theta}\cos{\phi} \cr
x_2 &=& \sin{\theta}\sin{\phi}\cr
x_3 &=& \cos{\theta}\;.
\eea
One can then explicitly evaluate the sets of identities 
\bea
K_i^aK_i^b&=&\hat h^{ab}\cr
\epsilon_{ijk} x_i K_j^a K_k^b&=&\hat\omega^{ab} =\frac{\epsilon^{ab}}{\sqrt{\hat h}} \cr
 K_i^ah_{ab}K^b_j&=&\delta_{ij}-x_ix_j\cr
K_i^a\d_a K_i^b&=&\frac{1}{\sqrt{\hat h}}\d^b\sqrt{\hat h}\;.
\eea
Further identities that were used for calculations in the main body of this paper include
\bea
&& x_i \d^a K_i^b=\hat\omega^{ab}\cr
&&\epsilon_{ijk}\d_a K_i^b x_j K_k^a=0\cr
&& \epsilon_{ijk}\d_a K_i^b K_j^c K_k^a\times (sym. b\leftrightarrow c)=0\cr
&& (\d_a x_i)K_j^a=\epsilon_{ijk}x_k\;.
\eea
From the last relation we also obtain 
\bea
 (\d_a x_i)K_i^a&=&0\cr
 \epsilon_{ijk}(\d_a x_i)K_j^a x_k&=&2\cr
 \epsilon_{ijk}(\d_a x_i) K_j^a K_k^b&=&0\;.
\eea

The 2d gamma-matrices in spherical coordinates can then be obtained with the knowledge of the vielbeins
\be
e^i_a = \textrm{diag}(1,\sin{\theta}) 
\ee
and the fact that in a Cartesian coordinate frame the 2d gamma-matrices are $\Gamma_i = \sigma_i$, with $i=1,2$ and $\sigma_i$ the usual Pauli matrices. The chirality matrix is given by $\hat \gamma _3 = - i \sigma_1 \sigma_2 =  \sigma_3 = \Gamma_3 $. Then
\be
{(\Gamma^\theta)^\alpha}_\beta ={(\sigma_1)^\alpha}_\beta\qquad \textrm{and}\qquad{(\Gamma^\phi)^\alpha}_\beta =\frac{1}{\sin{\theta}}{(\sigma_2)^\alpha}_\beta\;.
\ee
In going between Cartesian and spherical expressions on the sphere we  make use of the following unitary rotation matrix
\be
S=\sqrt{i}\begin{pmatrix}
-\sin{\frac{\theta}{2}} e^{i\phi/2}& -i\cos{\frac{\theta}{2}} e^{i\phi/2}\\
\cos{\frac{\theta}{2}} e^{-i\phi/2}&-i\sin{\frac{\theta}{2}} e^{-i\phi/2}
\end{pmatrix}\;.
\ee
One can then show that 
\bea\label{SKx}
{(S \Gamma^a S^{-1})^\a}_\b&=&-K_i^a{(\ts_i)^\a}_\b\cr
{(S \Gamma_3 S^{-1})^\a}_\b&=&-x_i^a{(\ts_i)^\a}_\b\cr
{(S P_- S^{-1})^\a}_\b&=& \delta^\alpha_\beta+x_i^a{(\ts_i)^\a}_\b\;,
\eea
where $P_\pm = \frac{1}{2}(1 \pm \Gamma_3)$ are projectors for gamma matrices in 2d.

\section{Gamma matrix relations and conventions}\label{AppB}

We first define some conventions for spinors in 2+1d. We will follow the standard ABJM notation of 
\cite{Benna:2008zy}, so that for worldvolume metric 
$\eta^{\mu\nu} = \text{diag}(-1,+1,+1)$ with $\mu = 0,1,2$ one uses Dirac matrices $\gamma^\mu = (i\sigma^2, \sigma^1,\sigma^3)$ satisfying $\gamma^\mu \gamma^\nu = \eta^{\mu\nu}+\epsilon^{\mu\nu\lambda}\gamma_\lambda$. 
For completeness, the fermionic indices, which we will denote with a hat to avoid confusion with other indices, are raised and lowered as 
$\theta^{\hat a} = \epsilon^{\hat a \hat b}\theta_{\hat a\hat b}$ and $\theta_{\hat a} = \epsilon_{\hat a\hat b}\theta^{\hat b}$, with $\epsilon^{12} = -\epsilon_{12}=1$, so that $\epsilon^{\hat a\hat b}\epsilon_{\hat b \hat c} = -\delta^{\hat a}_{\hat c}$. 
Note that lowering the spinor indices on the $\gamma$'s makes them symmetric $\gamma^\mu_{\hat a\hat b} = (-\one, -\sigma^3, \sigma^1)$. In terms of notation that will mean that scalar fermion quantities will imply an index contraction as per the `SW-NE' rule $\psi^2= \psi \psi = \psi_{\hat a}\psi^{\hat a}$.

It can be checked that if one lowers the indices on ${(\ts_i)^\a}_\b$, one gets a symmetric matrix, $(\ts_i)_{\a\b}=(\ts_i)_{\b\a}$. Then the same  also applies for ${(\Gamma_a)^\a}_\b$ and ${(\Gamma_3)^\a}_\b$, 
\ie $(\Gamma_a)_{\a\b}=(\Gamma_a)_{\b\a}$ and $(\Gamma_3)_{\a\b}=(\Gamma_3)_{\b\a}$.
Then it can be easily shown that 
\be
{(\ts_i)^\a}_\b={(\ts_i)_\b}^\a\equiv \epsilon_{\b\b'}\epsilon^{\b\b'}{(\ts_i)^{\b'}}_{\a'}
\ee
and thus also 
\be
{(\sigma_i)^\a}_\b={(\sigma_i)_\b}^\a\;.
\ee
Similarly, one can also prove (using the reality condition for S) that
\be
{(S\sigma_i S^{-1})^\a}_\b={(S\sigma_i S^{-1})_\b}^\a
\ee
and hence
\bea
{(S\Gamma_a S^{-1})^\a}_\b&=&{(S\Gamma_a S^{-1})_\b}^\a\cr
{(S\Gamma_3 S^{-1})^\a}_\b&=&{(S\Gamma_3 S^{-1})_\b}^\a\;.
\eea
From the above it would seem that one does not need to remember the matrix (horizontal) order in the indices, but that is not so since there is one exception: 
\be
{\delta_\b}^\a\equiv \epsilon_{\a\a'}\epsilon^{\b\b'}{\delta^{\a'}}_{\b'}=-{\delta^\a}_\b\;,
\ee
which means that 
\be
G_\a G^{\dagger\b}={J_\a}^\b\rightarrow\frac{N}{2}{(x_m\ts_m +1)_\a}^\b
=\frac{N}{2}{(x_m\ts_m-1)^\b}_\a\;.
\ee

The following trace identities are also useful
\bea
&&\Tr[\ts_i\ts_j]=2\delta_{ij}\cr
&&\Tr[\ts_i\ts_j\ts_k]=-2\epsilon_{ijk}\cr
&&\Tr[\ts_i\ts_j\ts_k\ts_l]=2(\delta_{ij}\delta_{kl}+\delta_{il}\delta_{jk}-\delta_{ik}\delta_{jl})\cr
&&\Tr[\ts_i\ts_j\ts_k\ts_l\ts_m]=-2i(\delta_{ij}\epsilon_{klm}+\delta_{lm}\epsilon_{ijk}+\delta_{kl}\epsilon_{mij}
-\delta_{km}\epsilon_{ijl})\label{traces}\;.
\eea

\end{appendix}

\bibliographystyle{utphys}
\bibliography{abfuz2}

\end{document}